\documentclass{emulateapj}
\usepackage{apjfonts}
\usepackage{graphicx}
\usepackage{xspace}

\newcommand{\Msun}      {\mbox{$\rm\,M_{\mathord\odot}$}}

\begin{document}

\lefthead{Multiwavelength Observations of Swift~J1753.5--0127}
\righthead{Tomsick et al.}

\submitted{Accepted by ApJ}

\def\lsim{\mathrel{\lower .85ex\hbox{\rlap{$\sim$}\raise
.95ex\hbox{$<$} }}}
\def\gsim{\mathrel{\lower .80ex\hbox{\rlap{$\sim$}\raise
.90ex\hbox{$>$} }}}

\title{The Accreting Black Hole Swift~J1753.5--0127 from Radio to Hard X-Ray}

\author{John A. Tomsick\altaffilmark{1}, Farid Rahoui\altaffilmark{2,3}, Mari Kolehmainen\altaffilmark{4}, James Miller-Jones\altaffilmark{5}, Felix F\"{u}rst\altaffilmark{6}, Kazutaka Yamaoka\altaffilmark{7,8}, Hiroshi Akitaya\altaffilmark{9}, St\'{e}phane Corbel\altaffilmark{10,11}, Mickael Coriat\altaffilmark{12}, Chris Done\altaffilmark{13}, Poshak Gandhi\altaffilmark{14}, Fiona A. Harrison\altaffilmark{6}, Kuiyun Huang\altaffilmark{15}, Philip Kaaret\altaffilmark{16}, Emrah Kalemci\altaffilmark{17}, Yuka Kanda\altaffilmark{18}, Simone Migliari\altaffilmark{19}, Jon M. Miller\altaffilmark{20}, Yuki Moritani\altaffilmark{8,21}, Daniel Stern\altaffilmark{22}, Makoto Uemura\altaffilmark{8}, Yuji Urata\altaffilmark{23}}

\altaffiltext{1}{Space Sciences Laboratory, 7 Gauss Way, University of California, Berkeley, CA 94720-7450, USA}

\altaffiltext{2}{European Southern Observatory, Karl Schwarzschild-Strasse 2, 85748 Garching bei Munchen, Germany}

\altaffiltext{3}{Department of Astronomy, Harvard University, 60 Garden Street, Cambridge, MA 02138, USA}

\altaffiltext{4}{Astrophysics, Department of Physics, University of Oxford, Keble Road, Oxford OX1 3RH, UK}

\altaffiltext{5}{International Centre for Radio Astronomy Research, Curtin University, GPO Box U1987, Perth, WA 6845, Australia}

\altaffiltext{6}{California Institute of Technology, 1200 East California Boulevard, Pasadena, CA 91125, USA}

\altaffiltext{7}{Solar-Terrestrial Environment Laboratory, Nagoya University, Furocho, Chikusa-ku, Nagoya, Aichi 464-8601, Japan}

\altaffiltext{8}{Division of Particle and Astrophysical Science, Department of Physics, Nagoya University, Furo-cho, Chikusa-ku, Nagoya, Aichi 464-8602, Japan}

\altaffiltext{9}{Hiroshima Astrophysical Science Center, Hiroshima University, Kagamiyama, Higashi-Hiroshima, Hiroshima 739-8526, Japan}

\altaffiltext{10}{Laboratoire AIM (CEA/IRFU - CNRS/INSU - Universit\'{e} Paris Diderot), CEA DSM/IRFU/SAp, F-91191 Gif-sur-Yvette, France}

\altaffiltext{11}{Station de Radioastronomie de Nan\c{c}ay, Observatoire de Paris, PSL Research University, CNRS, Univ. Orl\'{e}ans, OSUC, 18330 Nan\c{c}ay, France}

\altaffiltext{12}{Institut de Recherche en Astrophysique et Plan\'{e}tologie (IRAP), 9 Avenue du Colonel Roche, 31028 Toulouse Cedex 4, France}

\altaffiltext{13}{Department of Physics, University of Durham, South Road, Durham DH1 3LE, UK}

\altaffiltext{14}{School of Physics \& Astronomy, University of Southampton, Highfield, Southampton SO17 1BJ, UK}

\altaffiltext{15}{Department of Mathematics and Science, National Taiwan Normal University, Lin-kou District, New Taipei City 24449, Taiwan}

\altaffiltext{16}{Department of Physics and Astronomy, University of Iowa, Van Allen Hall, Iowa City, IA 52242, USA}

\altaffiltext{17}{Sabanci University, Orhanli-Tuzla, Istanbul, 34956, Turkey}

\altaffiltext{18}{Department of Physical Science, Hiroshima University, Kagamiyama, Higashi-Hiroshima 739-8526, Japan}

\altaffiltext{19}{European Space Astronomy Centre, Apartado/P.O. Box 78, Villanueva de la Canada, E-28691 Madrid, Spain}

\altaffiltext{20}{Department of Astronomy, University of Michigan, 500 Church Street, Ann Arbor, MI 48109-1042, USA}

\altaffiltext{21}{Kavli Institute for the Physics and Mathematics of the Universe (WPI), The University of Tokyo, 5-1-5, Kashiwanoha, Kashiwa, 277-8583, Japan}

\altaffiltext{22}{Jet Propulsion Laboratory, California Institute of Technology, 4800 Oak Grove Drive, Pasadena, CA 91109, USA}

\altaffiltext{23}{Institute of Astronomy, National Central University, Chung-Li 32054, Taiwan}

\begin{abstract}

We report on multi-wavelength measurements of the accreting black hole Swift~J1753.5--0127 
in the hard state at low luminosity ($L\sim 2.7\times 10^{36}$\,erg\,s$^{-1}$ assuming a 
distance of $d = 3$\,kpc) in 2014 April.  The radio emission is optically thick synchrotron, 
presumably from a compact jet.  We take advantage of the low extinction ($E(B-V)=0.45$ 
from earlier work) and model the near-IR to UV emission with a multi-temperature disk model.  
Assuming a black hole mass of $M_{\rm BH} = 5$\Msun~and a system inclination of $i = 40^{\circ}$, 
the fits imply an inner radius for the disk of $R_{\rm in}/R_{g} > 212$\,$d_{3}$\,($M_{BH}$/5\Msun)$^{-1}$, 
where $R_{g}$ is the gravitational radius of the black hole, and $d_{3}$ is the distance to the
source in units of 3\,kpc.  The outer radius is $R_{\rm out}/R_{g} = 90,000$\,$d_{3}$\,($M_{BH}$/5\Msun)$^{-1}$, 
which corresponds to $6.6\times 10^{10}$\,$d_{3}$\,cm, consistent with the expected size of 
the disk given previous measurements of the size of the companion's Roche lobe.
The 0.5--240\,keV energy spectrum measured by {\em Swift}/XRT, {\em Suzaku} 
(XIS, PIN, and GSO), and {\em NuSTAR} is relatively well characterized by an absorbed 
power-law with a photon index of $\Gamma = 1.722\pm 0.003$ (90\% confidence error), but a 
significant improvement is seen when a second continuum component is added.  Reflection is 
a possibility, but no iron line is detected, implying a low iron abundance.  We are able 
to fit the entire (radio to 240\,keV) spectral energy distribution (SED) with a multi-temperature 
disk component, a Comptonization component, and a broken power-law, representing the 
emission from the compact jet.  The broken power-law cannot significantly contribute to 
the soft X-ray emission, and this may be related to why Swift~J1753.5--0127 is an outlier 
in the radio/X-ray correlation.  The broken power-law (i.e., the jet) might dominate 
above 20\,keV, which would constrain the break frequency to be between $2.4\times 10^{10}$\,Hz 
and $3.6\times 10^{12}$\,Hz.  Although the fits to the full SED do not include significant 
thermal emission in the X-ray band, previous observations have consistently seen such a 
component, and we find that there is evidence at the 3.1-$\sigma$ level for a disk-blackbody 
component with a temperature of $kT_{\rm in} = 150^{+30}_{-20}$\,eV and an inner radius of 
5--14\,$R_{g}$.  If this component is real, it might imply the presence of an inner optically 
thick accretion disk in addition to the strongly truncated ($R_{\rm in} > 212$\,$R_{g}$) disk.
We also perform X-ray timing analysis, and the power spectrum is dominated by a Lorentzian 
component with $\nu_{\rm max} = 0.110\pm 0.003$\,Hz and $\nu_{\rm max} = 0.16\pm 0.04$\,Hz as 
measured by XIS and XRT, respectively.  

\end{abstract}


\section{Introduction}

Most accreting stellar-mass black holes in binary systems exhibit large changes in 
luminosity over time, ranging from a substantial fraction of the Eddington limit 
($L_{\rm Edd}$) to $\sim$$10^{-8}$ or $\sim$$10^{-9}$\,$L_{\rm Edd}$.  In addition to 
changes in luminosity, these systems show other observational changes, including 
transitions between distinct spectral states that are similar from system-to-system 
\citep[e.g.,][]{mr06,belloni10}.  The thermal dominant (or soft) state has a strong 
thermal component from an optically thick accretion disk in the X-ray spectrum.  In 
the hard state on the other hand, this component contributes a lower fraction of the 
flux in the X-ray band.  The drop in flux is partly due to a decrease in the temperature 
of the component \citep{kalemci04}, moving its peak into the ultraviolet where it is 
difficult to measure due to interstellar absorption.  While the soft thermal X-ray 
emission weakens, there is a strong increase in the hard X-rays, and the X-ray spectrum 
in the hard state is dominated by a power-law, which often has an exponential cutoff 
above 50--100\,keV \citep{grove98,gm14}.  

Accreting black holes also emit in the radio band when they are in the hard state, 
and this is due to a powerful compact jet \citep{corbel00,fender01}.  At radio 
frequencies, the spectrum is dominated by a partially self-absorbed synchrotron
component that has a flat or rising spectrum ($F_{\nu}\propto \nu^{\alpha}$, where $\alpha\gsim 0$).  
The jet spectrum changes slope above the break frequency, $\nu_{\rm break}$, becoming
steeper because the frequency is sufficiently high that the entire jet is optically
thin.  In some cases, the measurement of $\nu_{\rm break}$ has been constrained to be in the 
infrared (IR) to optical \citep{cf02,gandhi11,rahoui11,russell13a,russell13b,russell14}, but 
its measurement can be complicated because of the other emission components (e.g., 
from the accretion disk or the optical companion) and also because the jet spectrum 
is likely significantly more complicated than a simple broken power-law 
\citep{mnw05,migliari07}.  

In the hard state, it is clear that there is a strong connection between the X-ray
and radio emission.  The fluxes in the two bands are correlated 
\citep{corbel00,corbel03,gfp03,ckk08,corbel13,gallo14}, and while early studies 
suggested that all black hole sources might lie on the same correlation line, 
observations of more systems have shown that this is not the case
\citep{jonker10,coriat11}.  A current topic of debate is whether all sources lie
on two correlation lines, one track for standard sources and one for outliers, or 
if there is a continuum of different tracks \citep{coriat11,corbel13,gallo14}.  
Another topic is how much, if any, of the X-ray emission originates in the jet.  
While the most typical hard state spectrum with an exponential cutoff is well 
described by thermal Comptonization, and it has been argued that it is unlikely that 
this emission is due to synchrotron emission from a jet \citep{zdziarski03}, some
black hole spectra appear to have multiple high-energy continuum components 
\citep{joinet07,rodriguez08b,bouchet09,droulans10,russell10}, and a jet origin is 
not ruled-out. In fact, Cygnus~X-1 often shows two high-energy components in the 
hard state, including an MeV component \citep{mcconnell00,rahoui11,zls12}, and the 
detection of strong polarization at $>$400\,keV favors a synchrotron origin 
\citep{laurent11,jourdain12}.

The fact that a jet is present in the hard state and that there is some connection
between the disk and the jet leads to the question of what we know about the disk 
properties.  The main question regarding the optically thick disk concerns the
location of the inner radius ($R_{\rm in}$).  One idea is that the black hole states 
are essentially determined by the mass accretion rate and $R_{\rm in}$ \citep{emn97}, 
with sources entering the hard state because of an increase in $R_{\rm in}$.  However, 
X-ray observations of sources in the bright hard state seem to contradict this since 
relativistically smeared reflection components are seen from some systems that imply 
that the disk remains close to the innermost stable circular orbit 
\citep[ISCO;][]{blum09,reis11,miller12,fabian12}.  In addition, thermal component
modeling has led to similar conclusions \citep{rfm10}.  While photon pile-up in CCD 
spectra has sparked some debate about iron line results \citep{miller06a,dd10,miller10}, 
more recent observations with the {\em Nuclear Spectroscopic Telescope Array (NuSTAR)} 
confirm strongly broadened and skewed iron lines in the bright hard state for GRS~1915+105 
\citep{miller13}, GRS~1739--278 \citep{miller15}, GX~339--4 \citep{fuerst15}, 
and Cygnus~X-1 \citep{parker15}.  For the case of GRS~1739--278, the luminosity 
is $\sim$5\%\,$L_{\rm Edd}$, and the inferred inner radius is $<$12\,$R_{g}$ \citep{miller15}, 
where $R_{g} = GM_{\rm BH}/c^{2}$ and $G$ is the gravitational constant, $M_{\rm BH}$ is the 
black hole mass, and $c$ is the speed of light.  Significantly truncated disks have been 
reported for the hard state at intermediate and low luminosities using reflection 
component modeling \citep{tomsick09c,shidatsu11,plant15} and also by modeling the 
thermal component from the optically thick disk \citep{gdp08,cabanac09}.  

To investigate questions related to the accretion geometry and the relationship
between the disk and the jet, we performed multi-wavelength observations of the
accreting black hole Swift~J1753.5--0127 in the hard state.  This system was 
first discovered in outburst in 2005 \citep{palmer05}, and it is very unusual 
in that it has been bright in X-rays for almost a decade.  The optical light
curve shows a 3.2 hour modulation, which has been interpreted as a superhump
period (a modulation due to tidal stresses on a precessing, elliptical 
accretion disk), suggesting that the orbital period is somewhat smaller than 
this \citep{zurita08}.  From radial velocity measurements, \cite{neustroev14} 
find a 2.85 hour signal, which is likely the true orbital period.  Thus, 
Swift~J1753.5--0127 has one of the shortest orbital periods of any known
black hole binary.  Although the mass of the black hole in Swift~J1753.5--0127
is still debated since there has not been an opportunity to obtain a radial
velocity measurement for the companion star with the system in quiescence, 
\cite{neustroev14} argue that the mass is relatively low, $M_{\rm BH} < 5$\Msun, 
and we adopt a black hole mass of 5\Msun~for calculations in this paper.

Swift~J1753.5--0127 is also unusual in that it has a low level of extinction,
due in part to it being somewhat out of the plane with Galactic coordinates 
of $l = 24.9^{\circ}$ and $b = +12.2^{\circ}$.  \cite{froning14} obtained UV 
measurements showing that $E(B-V) = 0.45$, and we confirm this value in a 
companion paper (Rahoui et al., submitted to ApJ).  It is not entirely clear whether
the system is relatively nearby or in the Galactic halo as there is a large
range of possible distances, $d = 1$--10\,kpc \citep{cadollebel07,zurita08,froning14}.
\cite{froning14} provide evidence that the UV emission from Swift~J1753.5--0127 
in the hard state comes from an accretion disk, and they calculate distance upper 
limits that depend on $M_{\rm BH}$, assuming that the mass accretion rate is less 
than 5\% $L_{\rm Edd}$.  For $M_{\rm BH} = 5$\Msun, the upper limit is 2.8--3.7\,kpc, 
depending on the inclination of the system, and we use a fiducial distance of 
3\,kpc for the calculations in this paper.

Swift~J1753.5--0127 has been extensively observed in the radio band, and is one
of the clearest examples of a source that is an outlier in the radio/X-ray
correlation plot \citep{soleri10,corbel13}.  The location on the plot depends
on the assumed distance, and the previous work has assumed a source distance
of 8\,kpc.  While we are adopting a significantly smaller distance, \cite{soleri10}
considered how distance affects the the radio underluminosity, which is a measure
of how far a source is from the standard correlation.  \cite{soleri10} show that 
a smaller distance moves the source farther from the standard correlation
\citep[see also][]{jonker04a}.  Thus, the fact that recent work suggests that 
Swift~J1753.5--0127 is closer than early estimates only strengthens the conclusion 
that the source is an outlier.

\begin{figure}
\hspace{-0.3cm}
\includegraphics[scale=0.52]{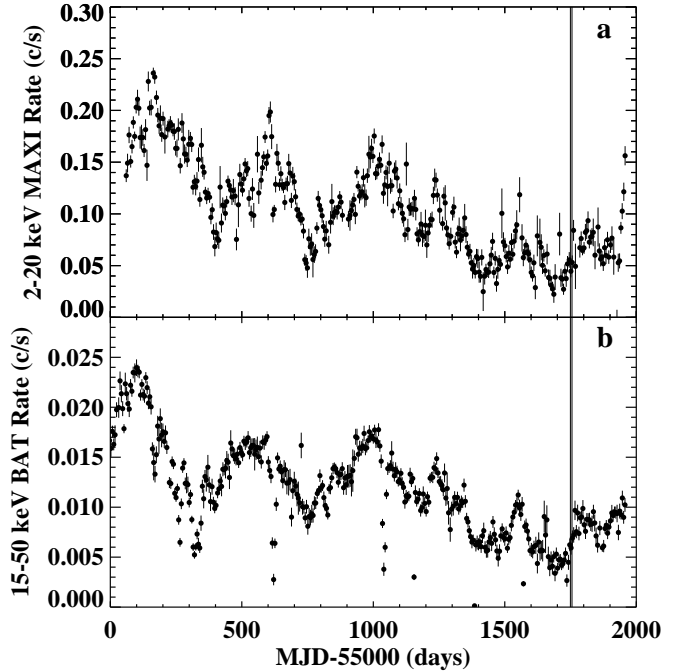}
\vspace{0.2cm}
\caption{\small {\em MAXI} {\it (a)} and {\em Swift}/BAT {\it (b)} light curves 
in the 2--20\,keV and 15--50\,keV bands for Swift~J1753.5--0127 between mid-2009 
and mid-2014.  The vertical lines mark the start and stop times of the observations 
used in this work (2014 April 2--8).\label{fig:lc}}
\end{figure}

For this work, we have carried out a large campaign to observe Swift~J1753.5--0127
in the hard state with radio, near-IR, optical, UV, and X-ray observations as
described in Section 2.  In the X-ray, data were obtained with {\em NuSTAR}, 
{\em Suzaku}, and {\em Swift}/XRT.  The observations occurred when the flux level 
was close to the minimum brightness this source has had in the $\sim$10 years since 
its discovery (see Figure~\ref{fig:lc}).  The low flux level (and presumably mass 
accretion rate) may cause changes in the properties of the accretion disk or jet 
compared to previous observations at higher flux levels.  In Section 3, we perform 
spectral analysis for the different energy ranges (radio, near-IR to UV, and X-ray) 
separately and then also as a combined radio to X-ray Spectral Energy Distribution 
(SED).  We also produce an X-ray power spectrum for timing analysis.  We discuss 
the results in Section 4, and then provide conclusions in Section 5.

\section{Observations and Data Reduction}

\begin{figure}
\hspace{-0.3cm}
\includegraphics[scale=0.52]{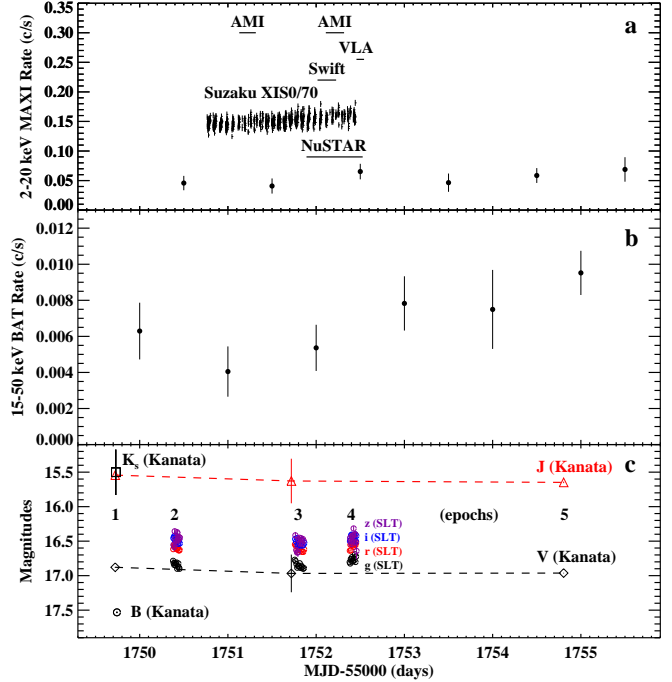}
\vspace{0.2cm}
\caption{\small {\it (a)} The filled circles show the {\em MAXI} light curve
over the time of the observations (2014 April 2--8). The times of the AMI, VLA, 
{\em Swift}/XRT, {\em Suzaku}, and {\em NuSTAR} observations are indicated.  
Also, the 1--12\,keV light curve for {\em Suzaku}/XIS0 is shown (the actual 
count rate divided by 70).  {\it (b)} The {\em Swift}/BAT light curve over 
the time of the observations.  {\it (c)} The optical and near-IR magnitudes 
measured at the Kanata and SLT telescopes.\label{fig:lc_zoom}}
\end{figure}

The observations that we obtained in 2014 April are listed in Table~\ref{tab:obs}, 
and more details about the observation times are shown in Figure~\ref{fig:lc_zoom}.  
The X-ray flux was rising very slowly during the observation, and this is seen 
especially clearly in the {\em Suzaku}/XIS light curve (Figure~\ref{fig:lc_zoom}).  
We provide more details about the observatories used and how the data were processed 
in the following.

\subsection{Radio}

We observed Swift~J1753.5--0127 with the Karl G. Jansky Very Large Array (VLA) 
on 2014 April 5 (MJD 56752) from 11:00--13:00 UT with the array in its 
most-extended A-configuration.  We split the observing time between the 4--8 
and 18--26-GHz observing bands.  In the lower 4--8\,GHz band, we split the 
available bandwidth into two 1024-MHz basebands, centered at 5.25 and 7.45\,GHz.  
Each baseband was split into eight 128-MHz spectral windows, each of which 
comprised sixty-four 2-MHz channels.  The higher-frequency 18--26\,GHz band 
was fully covered by four 2048-MHz basebands, each comprising sixteen 128-MHz 
sub-bands made up of sixty-four 2-MHz channels.  After accounting for calibration 
overheads, the total on-source integration times for Swift~J1753.5--0127 were 
25.3 minutes in the 4--8-GHz band and 29.1 minutes in the 18--26-GHz band.

The data were reduced using version 4.2.0 of the Common Astronomy Software Application 
\citep[CASA;][]{McMullin07}.  We applied a priori calibration to account for 
updated antenna positions and gain variations with changing elevation or 
correlator configuration, and corrected the 18--26-GHz data for opacity 
effects. We edited out any data affected by antenna shadowing before Hanning 
smoothing the data and removing any radio frequency interference.  At all 
frequencies we used 3C286 to calibrate the instrumental frequency response, 
and to set the amplitude scale according to the default Perley-Butler 2010 
coefficients implemented in the CASA task \textsc{setjy}.  We used J1743-0350 
as a secondary calibrator to determine the time-varying complex gains arising 
from both atmospheric and instrumental effects.

The calibrated data on Swift~J1753.5--0127 were averaged by a factor of four 
in frequency to reduce the raw data volume, and then imaged using Briggs 
weighting with a robust parameter of 1 to achieve the best compromise between 
sensitivity and sidelobe suppression.  When imaging, we used the multi-frequency 
synthesis algorithm as implemented in CASA's \textsc{clean} task, choosing two 
Taylor terms to account for the frequency dependence of source brightness.  
The source was clearly detected in all frequency bands, with an inverted 
spectrum ($\alpha>0$, where the flux density $S_{\nu}$ varies as a function 
of frequency $\nu$ as $S_{\nu}\propto\nu^{\alpha}$).  To better constrain the 
radio spectrum, we split each frequency band into four frequency bins (of 
width 1024\,MHz at 4--8\,GHz, and 2048\,MHz at 18--26\,GHz where the intrinsic 
sensitivity per unit frequency is lower).  We measured the source brightness 
in each frequency bin by fitting an elliptical Gaussian to the brightness 
distribution in the image plane.  Swift~J1753.5--0127 appeared unresolved at 
all frequencies.

Swift~J1753.5$-$0127 was also observed with the Mullard Radio Astronomy Observatory's 
Arcminute Microkelvin Imager (AMI) Large Array \citep{zwart08} on two occasions 
during the coordinated multi-wavelength campaign between 2014 April 4--5. These 
$\sim$4-hour observations were taken at the times given in Table~\ref{tab:obs} 
with a central frequency of 15.4~GHz.  The AMI Large Array consists of eight 
13-m dishes, with the full frequency band of 12--17.9~GHz being divided into eight 
0.75~GHz bandwidth channels. Channels 1--2 and 8 were ignored due to lower response 
in those frequency ranges and the expected high level of interference from satellites 
due to the low elevation of the source. The primary beam FWHM is $\approx$6 arcmin at 
16~GHz.

The AMI data were reduced using the semi-automated pipeline procedure described in 
\cite{staley13}, which uses the AMI software tool {\sc reduce} to automatically flag 
for interference, shadowing and hardware errors, calibrate the gain, and synthesize
the frequency channels to produce visibility data in uv-FITS format 
\citep[see][for more details]{staley13}.  However, the low elevation and the 
radio-quiet nature of the source resulted in high noise levels in the reduced images, 
and thus the two observations were concatenated to maximize the signal-to-noise. The 
concatenated dataset was then imaged in CASA, where the {\sc clean} task was used to 
produce the combined frequency image, and the flux density was measured by fitting a 
Gaussian model to the source in the radio map using the MIRIAD task {\sc imfit}. The 
error on the concatenated flux density was calculated as  
$\sigma=\sqrt{(0.05S_{\nu})^{2}+{\sigma}^{2}_{\rm fit}+{\sigma}^{2}_{\rm rms} }$ following 
\cite{ainsworth12}, with a 5\% absolute calibration error added to the fitting error 
${\sigma}_{\rm fit}$ calculated in {\sc imfit}. The source concatenated flux density was 
measured at $290\pm50~\mu$Jy.

\subsection{Ground-Based Optical and Near-IR}

Kanata is a 1.5 m telescope at the Higashi-Hiroshima Observatory.  Photometric 
observations were performed for this study on three nights (MJD 56749, 56751, and 56754)
with the $B$, $V$, $J$, and $K_{s}$ bands using the HONIR instrument \citep{sakimoto12,akitaya14}
attached to Kanata.  The individual frame exposure times were 75, 136, 120, and 60\,s in $B$, 
$V$, $J$, $K_{s}$-bands, respectively.  The data reduction was performed in the standard 
manner: the bias and dark images were subtracted from all images, and then the images were 
flat-fielded. The magnitudes of the object and comparison stars were measured using PSF 
photometry.  For the $B$, $V$, and $J$-band photometry, we used the comparison star located 
at R.A.=$17^{\rm h}53^{\rm m}25.\!^{\rm s}275$, 
Decl.=--$01^{\circ}27^{\prime}30.\!^{\prime\prime}05$ (J2000.0), which has magnitudes of 
$B = 17.62$, $V = 16.66$, and $J = 14.468$ \citep{zurita08,skrutskie06}.  For the $K_{s}$-band 
photometry, we used the comparison star at R.A.=$17^{\rm h}53^{\rm m}25.\!^{\rm s}853$,
Decl.=--$01^{\circ}26^{\prime}17.\!^{\prime\prime}00$ (J2000.0), for which $K_{s} = 11.132$ 
\citep{skrutskie06}.

We also conducted optical $g'$, $r'$, $i'$, and $z'$ band monitoring observations with the 
Lulin 41 cm Super-Light Telescope (SLT), which is located in Taiwan, on three nights in 
2014 April (see Table~\ref{tab:obs}).  Photometric images with 180\,s exposures were obtained 
using the U42 CCD camera. We performed the dark-subtraction and flat-fielding correction using 
the appropriate calibration data with the IRAF package. Photometric calibrations were made with 
the Pan-STARRS1 $3\pi$ catalogs \citep{magnier13,schlafly12,tonry12}. The DAOPHOT package was 
used to perform the aperture photometry of the multi-band images.

\subsection{Swift}

The {\em Swift} satellite \citep{gehrels04} includes two pointed instruments, the X-ray 
Telescope \citep[XRT;][]{burrows05} and the Ultra-Violet/Optical Telescope
\citep[UVOT;][]{roming05}, and we used data from both instruments from ObsID 00080730001
in this work.  We performed the XRT data reduction using HEASOFT v6.15.1 and the 2013 March 
version of the XRT calibration data base (CALDB), and made event lists using 
{\ttfamily xrtpipeline}.  The XRT instrument was in Windowed Timing mode to avoid photon 
pile-up.  For spectral analysis, we extracted photons from within $47^{\prime\prime}$ of the 
Swift~J1753.5--0127 position, and made a background spectrum from a region away from the 
source.  We measured an XRT source count rate of 7.7\,c/s in the 0.5--10\,keV band during 
the 2.4\,ks observation.  We used the appropriate response file from the CALDB 
(swxwt0to2s6\_20010101v015.rmf) and produced a new ancillary response file using 
{\ttfamily xrtmkarf} and the exposure map generated by {\ttfamily xrtpipeline}.  We 
binned the 0.5--10\,keV spectra so that each bin has a signal-to-noise of 10.

For UVOT, we obtained photometry in six filters ($v$, $b$, $u$, $uvw1$, $uvm2$, and
$uvw2$) during the observation.  For each filter, we produced an image using 
{\ttfamily uvotimsum} and made a source region with a radius of $5^{\prime\prime}$ and
a background region from a source-free region.  Then, we used {\ttfamily uvotsource}
to perform the photometry and calculate the magnitude and flux of Swift~J1753.5--0127
for each filter.  

{\em Swift} also includes the wide field of view (FoV) Burst Alert Telescope (BAT), 
and we use data from BAT to study the long-term 15--50\,keV flux (see Figures~\ref{fig:lc}
and \ref{fig:lc_zoom}).

\subsection{NuSTAR}

The {\em Nuclear Spectroscopic Telescope Array} \citep[{\em NuSTAR};][]{harrison13} 
consists of two co-aligned X-ray telescopes, FPMA and FPMB, sensitive between 3--79\,keV. 
To reduce the data, we used {\ttfamily nupipeline} v.1.3.1 as distributed with 
HEASOFT 6.15.1.  During our analysis, an updated version became available, but we 
carefully checked that it does not influence our results.  We extracted the source 
spectrum from a circular region with $90^{\prime\prime}$ radius centered on the J2000 
coordinates.  Due to the triggered readout of the detectors, pile-up is not a concern 
for {\em NuSTAR}.  The background was extracted from a circular region with a
$170^{\prime\prime}$ radius at the other end of the FoV. Small systematic 
changes of the background over the FoV can be neglected, as Swift~J1753.5--0127 is a 
factor of six brighter than the background, even at 70\,keV.  The spectrum includes 
data from two {\em NuSTAR} ObsIDs.  We reduced both ObsIDs separately and added the 
resulting spectra and response files using {\ttfamily addascaspec}. The resulting 
total exposure time is given in Table~\ref{tab:obs}.

\subsection{Suzaku}

For {\em Suzaku}, we used data from the X-ray Imaging Spectrometers 
\citep[XISs;][]{koyama07} and from the Hard X-ray Detector 
\citep[HXD;][]{takahashi:07a} PIN diode detector, and the HXD gadolinium 
silicate crystal detector (GSO).  The XIS has three CCD detectors
(XIS0, XIS1, and XIS3) that operate in the 0.4--12\,keV bandpass.  We
produced event lists for each detector using {\ttfamily aepipeline} and
merged the event lists taken in the 3$\times$3 and 5$\times$5 CCD editing
modes.  We ran {\ttfamily aeattcor2} and {\ttfamily xiscoord} on each
of the merged event files to update the attitude correction because this is 
important for the pile-up estimate, which we calculated using {\ttfamily pileest}.
We extracted source spectra using a $4^{\prime}$-radius circle with the inner 
$22^{\prime\prime}$ removed due to pile-up at a level of $>$4\% in the core
of the point spread function (PSF).  We extracted the background from a 
rectangular region near the edge of the active area of the detector.  
The XIS detectors were in 1/4 window mode for the observation, and part of
the source region falls off of the active region of the detector.  We 
accounted for this when determining the background scaling.  We used
{\ttfamily xisrmfgen} and {\ttfamily xissimarfgen} to produce response
matrices, and we combined the XIS0 and XIS3 spectra (the two front-illuminated 
CCD detectors) into a single file.

For HXD, we analyzed both PIN and GSO data using the Perl scripts {\tt hxdpinxbpi} 
and  {\tt hxdgsoxbpi}, respectively, after screening with the standard selection 
criteria.  These scripts produce deadtime-corrected source and background spectra 
automatically.  The non X-ray background model was taken from the FTP 
sites\footnote{ftp://legacy.gsfc.nasa.gov/suzaku/data/background/pinnxb\_ver2.2\_tuned/ and ftp://legacy.gsfc.nasa.gov/suzaku/data/background/gsonxb\_ver2.6/}, 
and cosmic X-ray background (CXB) was also subtracted based on previous 
{\em High Energy Astronomy Observatory (HEAO)} observations \citep{gruber99} for 
PIN. As an energy response, we used ae\_hxd\_pinxinome11\_20110601.rsp for PIN 
and ae\_hxd\_gsoxinom\_20100524.rsp with an additional correction file 
(ae\_hxd\_gsoxinom\_crab\_20100526.arf) for GSO.  The background count rate is
significantly higher than the source rate for GSO, but we still clearly detect
Swift~J1753.5--0127 at a rate of $0.740\pm 0.026$ c/s.

\section{Results}

\subsection{Energy Spectrum}

We performed all of the spectral fits using the XSPEC v12.8.2 software.  For the X-ray 
spectra, we used instrument response files produced using the HEASOFT software.  For the 
radio, ground-based optical and near-IR, and UVOT, we determined the flux for each data 
point and then used {\ttfamily flx2xsp} to produce spectral files and unitary response
matrices that can be read into XSPEC.  All spectral fits are performed by minimizing 
the $\chi^{2}$ statistic.

\subsubsection{Radio Spectrum}

\begin{figure}
\hspace{-0.5cm}
\includegraphics[scale=0.52]{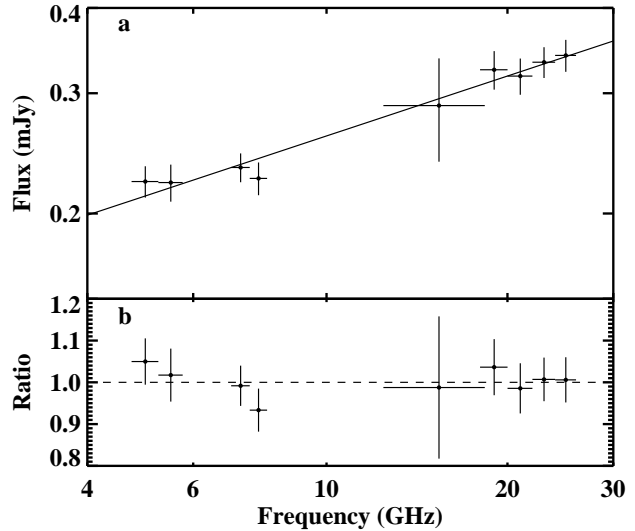}
\vspace{-1.5cm}
\caption{\small {\it (a)} Power-law fit to the radio spectrum with 
$\Gamma = 0.71\pm 0.05$ ($\alpha = 0.29\pm 0.05$).  The measurements
are from VLA (8 points) and AMI (1 point at 15.4\,GHz).
{\it (b)} Data-to-model ratio.\label{fig:spectrum_radio}}
\end{figure}

We fitted the radio points with a power-law model (see Figure~\ref{fig:spectrum_radio}), 
and this provides an acceptable fit with a reduced-$\chi^{2}$ ($\chi^{2}_{\nu}$) of 0.41 
for 7 degrees of freedom (dof).  The power-law photon index is $\Gamma = 0.71\pm 0.05$ 
(90\% confidence errors are given here and throughout the paper unless otherwise indicated),
and this corresponds to a spectral index of $\alpha = 1-\Gamma = 0.29\pm 0.05$
(as mentioned above, $\alpha$ is defined according to $S_{\nu}\propto\nu^{\alpha}$, where
$S_{\nu}$ is the flux density).  
We used the XSPEC model {\ttfamily pegpwrlw}, allowing for the power-law normalization 
to be defined as the flux density at 10\,GHz, and we obtain a measurement of 
$256\pm 8$\,$\mu$Jy at this frequency.

\subsubsection{Near-IR to UV Spectrum}

The times of the data taken for the near-IR to UV part of the spectrum from Kanata, SLT, 
and {\em Swift}/UVOT are shown in Figure~\ref{fig:lc_zoom}.  The ground-based (Kanata
and SLT) observations were taken in five epochs over six nights (see 
Tables~\ref{tab:obs_kanata} and \ref{tab:obs_slt} for the exact times of the exposures).  
As the source is variable from night-to-night and also on shorter time scales, we used 
measurements as close to each other in time as possible, while keeping the maximum wavelength 
coverage.  The {\em Swift} observation occurred between epochs 3 and 4, and we used the 
points from epoch 3 because the Kanata $V$- and $J$-band measurements occurred on the same 
night.  We also used the $K_{s}$-band measurement from epoch 1 because the statistical error 
bar is large enough to account for source variability.  We did not include the $B$-band
measurement because UVOT covered the same frequency, and the UVOT measurement was closer 
in time to the other observations.  For each SLT band, several epoch 3 measurements were 
made, and for the SED, we used the average value.  We estimated the uncertainty on these 
points by calculating the standard deviation of the measurements.

We fitted the near-IR to UV spectrum with a power-law model with extinction.  The 
XSPEC extinction model, {\ttfamily redden}, is based on the \cite{ccm89} relationship.  
The fit is poor ($\chi^{2}_{\nu} = 10.8$ for 10 dof), strongly over-predicting the 
$K_{s}$-band point.  The photon index for the power-law is $\Gamma = 0.2\pm 0.2$, but 
we suspect that this is not physically meaningful.  A somewhat better fit (although 
still far from being formally acceptable) is obtained by replacing the power-law with 
a blackbody (specifically {\ttfamily bbodyrad}), and this model is a much better match 
to the spectral slope in the near-IR.  If $E(B-V)$ is left as a free parameter, 
$\chi^{2}_{\nu} = 8.0$ for 10 dof, and we find $E(B-V) = 0.54\pm 0.06$, $kT = 1.9\pm 0.3$\,eV, 
and a normalization of $R_{\rm km}/d_{10} = (1.0\pm 0.1)\times 10^{12}$, where $R_{\rm km}$ 
is the size of the blackbody in units of kilometers, and $d_{10}$ is the distance to the 
source in units of 10\,kpc.  Fixing $E(B-V)$ to 0.45 gives $\chi^{2}_{\nu} = 7.8$ for 11 dof 
(essentially the same quality as the fit with the extinction parameter free), 
$kT = 1.51\pm 0.05$\,eV, and $R_{\rm km}/d_{10} = (1.2\pm 0.1)\times 10^{12}$.  

\begin{figure}
\hspace{-0.5cm}
\includegraphics[scale=0.52]{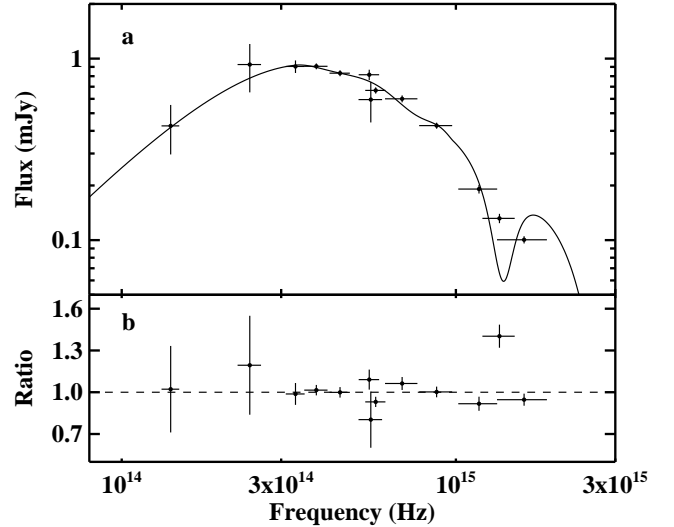}
\vspace{-1.5cm}
\caption{\small {\it (a)} Fit to the Kanata, SLT, and UVOT spectra with a 
multi-temperature disk model with outer radius as a free parameter.  The 
points are not dereddened, and the model assumes $E(B-V) = 0.45$.
{\it (b)} Data-to-model ratio.
\label{fig:spectrum_uv2ir}}
\end{figure}

The fit is worse with a multi-temperature disk-blackbody {\ttfamily diskbb} model 
($\chi^{2}_{\nu} = 13.4$ for 11 dof); however, a significant improvement is obtained 
if the outer edge of the disk is left as a free parameter.  We implemented this by
using the {\ttfamily diskir} model \citep{gdp08,gdp09}.  We turned off the thermalization
in the outer disk ($f_{\rm out} = 0$), and we set the Compton fraction ($L_{c}/L_{d}$) to zero.  
This model gives $\chi^{2}_{\nu} = 3.6$ for 10 dof, and the near-IR to UV spectrum is shown 
fitted with this model in Figure~\ref{fig:spectrum_uv2ir}.  For the parameters, we obtain 
$kT_{\rm in} = 5^{+2}_{-1}$\,eV for the temperature of the inner disk and a value of 
$1.29^{+0.26}_{-0.23}$ for $\log{r_{\rm out}}$, where $r_{\rm out} = R_{\rm out}/R_{\rm in}$, 
and $R_{\rm in}$ and $R_{\rm out}$ are, respectively, the inner and outer radii of the 
optically thick accretion disk.  The {\ttfamily diskir} normalization, which has the 
same meaning as the {\ttfamily diskbb} normalization  
($N_{\rm diskbb} = (R_{\rm in,km}/d_{10})^{2}/\cos{i}$, where $R_{\rm in,km}$ is the inner 
radius in units of kilometers, $d_{10}$ is the distance to the source in units of 
10\,kpc, and $i$ is the inclination of the disk) is 
$N_{\rm diskbb} = (9^{+11}_{-5})\times 10^{9}$.  Here, we simply note that this implies a 
very large inner disk radius.  We consider the implications below in detail after using 
the same model as a component in fitting the full SED.

None of the fits described above are formally acceptable, and there are a few
possible reasons for this.  Of course, the first possibility is that the spectrum
requires a more complex model than those we have tried.  Secondly, it is known that
there is significant variability in this part of the spectrum \citep{zurita08,neustroev14}, 
and this is also seen in Figure~\ref{fig:lc_zoom}.  Finally, the largest residuals 
(see Figure~\ref{fig:spectrum_uv2ir}) are in the UV where the extinction changes 
rapidly.  Uncertainties in the extinction law and the calibration of the broad UVOT 
photometric bins could also lead to the large residuals in this part of the spectrum.

\subsubsection{X-ray Spectrum}

\begin{figure}
\hspace{-0.5cm}
\includegraphics[scale=0.52]{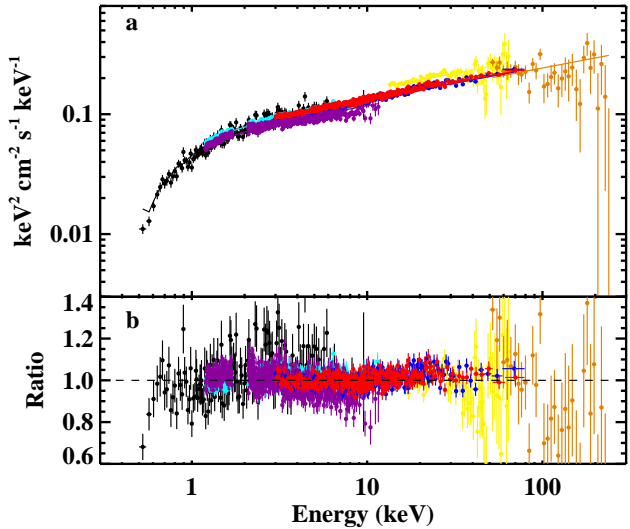}
\vspace{-1.5cm}
\caption{\small {\it (a)} Absorbed power-law fit to the Swift~J1753.5--0127
X-ray spectrum, including {\em Swift}/XRT (black), {\em Suzaku}/XIS03 (cyan), 
{\em Suzaku}/XIS1 (purple), {\em NuSTAR}/FPMA (blue), {\em NuSTAR}/FPMB 
(red), {\em Suzaku}/PIN (yellow), and {\em Suzaku}/GSO (orange).
{\it (b)} Data-to-model ratio.\label{fig:spectrum_x}}
\end{figure}

We performed a simultaneous fit to the spectra from all the X-ray instruments with an 
absorbed power-law model, allowing for different overall normalizations between 
instruments.  To account for absorption, we used the {\ttfamily tbabs} model with
\cite{wam00} abundances and \cite{vern96} cross-sections.  As shown in 
Table~\ref{tab:parameters_x}, the column density is
$N_{\rm H} = (2.01\pm 0.05)\times 10^{21}$\,cm$^{-2}$, the power-law photon index
is $\Gamma = 1.722\pm 0.003$, and this simple model provides a surprisingly good
fit with $\chi^{2}_{\nu} = 1.40$ for 2143 dof.  The residuals (see the data-to-model
ratio in Figure~\ref{fig:spectrum_x}b) do not show any evidence for an iron emission 
line as might be expected if there was a strong reflection component.  For a narrow 
line in the 6.4--7.1\,keV range, the 90\% confidence upper limit on the equivalent 
width is $<$5\,eV, and for a line with a width of 0.5\,keV, the upper limit on the 
equivalent width is $<$6\,eV.  The {\em Suzaku}/GSO shows a different slope above 
$\approx$80\,keV, and we added an exponential cutoff using the {\ttfamily highecut} 
model.  A cutoff with $E_{\rm cut} = 66^{+15}_{-10}$\,keV and $E_{\rm fold} = 218^{+151}_{-70}$\,keV 
provides a large improvement in the fit to the GSO data, but the overall $\chi^{2}_{\nu}$ 
only improves to 1.39 for 2141 dof.  

Previous work fitting X-ray spectra of Swift~J1753.5--0127 has often shown evidence 
for a thermal disk-blackbody component with an inner disk temperature of 
$kT_{\rm in} = 0.1$--0.4\,keV when the source is in the hard state
\citep{miller06b,hiemstra09,chiang10,reynolds10,cassatella12,kolehmainen14}.  Thus, 
we added a {\ttfamily diskbb} model to the power-law with an exponential cutoff, and 
the $\chi^{2}_{\nu}$ improves to 1.29 for 2139 dof (see Table~\ref{tab:parameters_x}).
While this represents a significant improvement (an F-test indicates that the significance
of the additional component is in excess of 12-$\sigma$), the temperature is much higher
and the normalization is much lower than previously seen.  Our value is 
$N_{\rm diskbb} = 3.8^{+1.5}_{-1.1}$ compared to values of $\gsim$1000 reported by
\cite{reynolds10} and \cite{cassatella12}.  A value of $N_{\rm diskbb} = 3.8$ would 
imply an unphysically small inner radius.  The equation for the inner radius in 
terms of the gravitational radius is
\begin{equation}
R_{\rm in}/R_{g} = (0.676~d_{10} f^{2} \sqrt{N_{\rm diskbb}})/((M_{\rm BH}/\Msun)\sqrt{\cos{i}})~~~~,
\end{equation}
where $f$ is the spectral hardening factor \citep{st95}.  For a distance of 3\,kpc, 
$M_{\rm BH}/\Msun = 5$, $f = 1.7$, which is a typical value \citep{st95}, and $i = 40^{\circ}$
based on the estimate of \cite{neustroev14}, we find $R_{\rm in}/R_{g} = 0.26$, which
puts the inner radius inside the event horizon.  

Figure~\ref{fig:spectrum_x} shows that there is a small deviation from the power-law 
in the hard X-ray band with the residuals increasing above 10\,keV and peaking near 
25\,keV.  Although there is no iron line, this could still be evidence for a weak 
reflection component or an additional continuum parameter.  Adding a reflection component 
to the power-law using the {\ttfamily reflionx} model \citep{rf05} provides a significant 
improvement in the fit to $\chi^{2}_{\nu} = 1.27$ for 2138 dof.  The reflection covering 
fraction (determined by calculating the ratio of the 0.001--1000\,keV unabsorbed flux in the
reflection component to the 0.1--1000\,keV unabsorbed flux in the direct component)
of $\Omega/2\pi = 0.2$ and the ionization (parameterized by $\xi = L/nR^{2}$, 
where $L$ is the luminosity of ionizing radiation, $n$ is the electron number density, 
and $R$ is the distance between the source of radiation and the reflecting material)
of $\xi < 5.3$\,erg\,cm\,s$^{-1}$ (see Table~\ref{tab:parameters_x}) would both be 
reasonable for a cool and truncated disk \citep[although we note that low covering 
fractions can also be explained by beaming emission away from the disk;][]{beloborodov99}.  
The iron abundance of $0.28\pm 0.08$ times solar is low but perhaps not unreasonably so.  
Adding a {\ttfamily diskbb} in addition to {\ttfamily reflionx} only provides a small 
improvement to the fit (to $\chi^{2}_{\nu} = 1.26$ for 2136 dof), and $N_{\rm diskbb}$ 
is even smaller than the previous value.  However, it is notable that adding the 
{\ttfamily diskbb} component causes the iron abundance to change to $0.47^{+0.21}_{-0.15}$
times solar.

\subsubsection{XRT Spectrum and the Possibility of a Thermal Component}

To investigate further on the question of why we do not see a physically reasonable
{\ttfamily diskbb} component while many previous studies of Swift~J1753.5--0127
in the hard state did, we fit the X-ray spectra individually.  Despite the short 
exposure time, the {\em Swift}/XRT spectrum provides the best information on this 
because it extends down to 0.5\,keV without strong instrumental features (we note 
that {\em Suzaku}/XIS also has sensitivity down at this energy, but the residuals 
indicate the presence of instrumental features).  A fit to the XRT spectrum with 
an absorbed power-law model gives $N_{\rm H} = (2.2\pm 0.2)\times 10^{21}$\,cm$^{-2}$, 
$\Gamma = 1.65\pm 0.03$, and $\chi^{2}_{\nu} = 1.27$ for 131 dof.  Adding a 
{\ttfamily diskbb} provides a significant improvement (to $\chi^{2}_{\nu} = 1.17$ 
for 129 dof), and an F-test indicates a significance of 99.8\% (3.1-$\sigma$) 
for the {\ttfamily diskbb} component.  The parameters for this fit are 
$N_{\rm H} = (5\pm 1)\times 10^{21}$\,cm$^{-2}$, $\Gamma = 1.76\pm 0.06$, 
$kT_{\rm in} = 130^{+20}_{-10}$\,eV, and $N_{\rm diskbb} = (1.6^{+3.0}_{-1.2})\times 10^{5}$.  

Although the column density is not known precisely, it is clear that it is lower than 
$\approx$$6\times 10^{21}$\,cm$^{-2}$.  The extinction value that we use in this paper
($E(B-V) = 0.45$) corresponds to $N_{\rm H} = 3.1\times 10^{21}$\,cm$^{-2}$ based on
the relationship derived in \cite{go09}.  Fixing the column density to this value
and fitting the XRT spectrum with a model consisting of a {\ttfamily diskbb} and
a power-law gives thermal parameter values of $kT_{\rm in} = 150^{+30}_{-20}$\,eV
and $N_{\rm diskbb} = (1.1^{+1.7}_{-0.5})\times 10^{4}$.  Thus, if we only had the 
{\em Swift}/XRT data, we would likely conclude that there is a physically reasonable 
thermal component.  The $kT_{\rm in} = 150$\,eV {\ttfamily diskbb} component that 
may be present in the {\em Swift}/XRT spectrum falls rapidly going to energies below 
soft X-rays and cannot explain the near-IR to UV emission that we see.  Thus, even
if it is real (and it may not be because it does not appear to be present when fitting
all the available data), it is not one of the dominant components in the overall SED, 
and we do not include it in the following as we build a model for fitting the full SED.  

\subsubsection{Near-IR, Optical, UV, and X-ray Spectrum}

Before fitting the full SED, we fit the near-IR to X-ray spectrum in order to determine
if it can be fit in a physically self-consistent manner.  As we found that the near-IR 
to UV spectrum requires a thermal model with the outer disk radius as a parameter, we 
start by fitting the spectrum with a {\ttfamily diskir} model.  While the fits above 
used a Compton fraction of zero (no Comptonization component), here we allow 
$L_{c}/L_{d}$ to be a free parameter, so that the model includes Comptonization by a 
thermal distribution of electrons with a temperature of $kT_{e}$, causing the model to 
extend into the X-ray.  Within {\ttfamily diskir}, Comptonization is implemented with
the {\ttfamily nthcomp} model \citep{zjm96,zds99}.  The physical scenario being considered 
is a near-IR to UV thermal component from a truncated optically thick accretion disk, 
providing seed photons to a Comptonization region with hot electrons.

The {\ttfamily diskir} model alone provides a reasonably good description of the 
spectrum, but it is not formally acceptable with $\chi^{2}_{\nu} = 1.38$ for 2152 dof 
(see Table~\ref{tab:parameters2}).  The fact that the thermal component acts as the 
seed photon distribution for the Comptonization emission leads to a somewhat higher 
value of $kT_{\rm in}$ ($12^{+8}_{-5}$\,eV compared to $5^{+2}_{-1}$\,eV found in 
Section 3.1.2) and a lower normalization, corresponding to a somewhat smaller disk 
inner radius.  The temperature of the Comptonizing electrons is constrained to be 
$>$60\,keV, and the Comptonizing fraction is $L_{c}/L_{d} = 4.2^{+2.4}_{-2.1}$.  

Adding a second continuum component provides a much improved fit, and approximately 
the same improvement is seen whether we add a power-law with an exponential cutoff 
or a reflection component (see Table~\ref{tab:parameters2}).  Also, both two-component 
models lead to very similar values for the thermal component with $kT_{\rm in}$ 
increasing to $29^{+17}_{-7}$\,eV in one case and $29\pm 5$\,eV in the other.  The 
values of $N_{\rm diskbb}$ decrease further, but they still imply a large disk truncation 
radius.

\begin{figure}
\hspace{-0.5cm}
\includegraphics[scale=0.52]{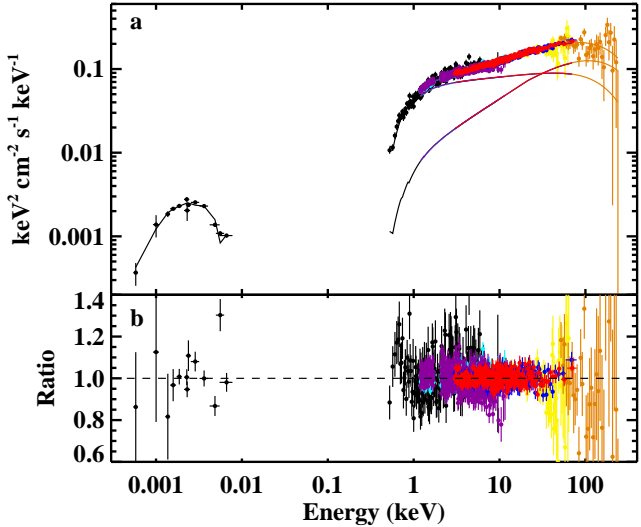}
\vspace{-1.5cm}
\caption{\small {\it (a)} Fit to the Swift~J1753.5--0127 spectrum, including the
data from the near-IR to the X-ray.  The model is {\ttfamily diskir} plus a power-law
with a high-energy cutoff.  The power-law is the harder of the two components.
{\it (b)} Data-to-model ratio.\label{fig:nir2x1}}
\end{figure}

The two-component models have very different implications for the properties of the 
Comptonization region.  The physical scenario we are considering in adding an extra 
power-law is that either this emission comes from the jet or that there is an
inhomogeneous or multi-phase Comptonization region \citep{makishima08,takahashi08,yamada13b}.  
When this component is added, as shown in Figure~\ref{fig:nir2x1}, its best fit parameters 
imply a very hard spectrum $\Gamma = 1.33^{+0.08}_{-0.25}$, and it dominates at high 
energies, so that the {\ttfamily diskir} Comptonization component can have much lower 
values of $kT_{e}$ (the constraint is $>$35\,keV) and $L_{c}/L_{d} = 0.77\pm 0.17$.  
A value of $L_{c}/L_{d}$ below 1.0 is unusual for the hard state, but this is due to
the fact that much of the hard X-ray flux is in the power-law component.

\begin{figure}
\hspace{-0.5cm}
\includegraphics[scale=0.52]{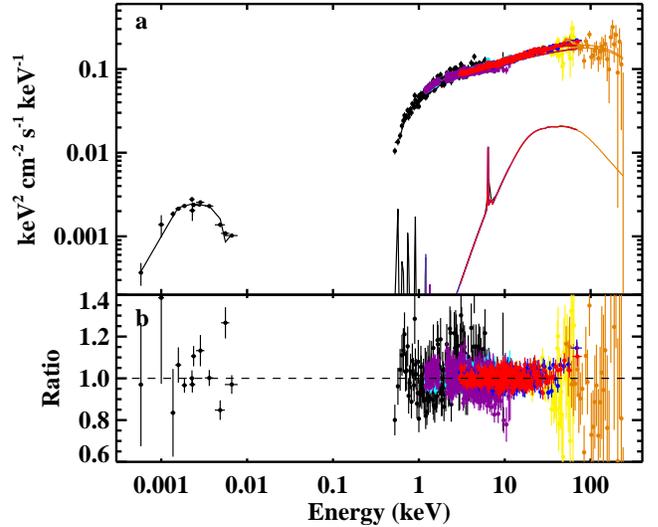}
\vspace{-1.5cm}
\caption{\small {\it (a)} Fit to the Swift~J1753.5--0127 spectrum, including the
data from the near-IR to the X-ray.  The model is {\ttfamily diskir} plus a 
reflection component.  {\it (b)} Data-to-model ratio.\label{fig:nir2x2}}
\end{figure}

On the other hand, when reflection is added, as shown in Figure~\ref{fig:nir2x2},
the physical scenario is that the {\ttfamily diskir} Comptonization component 
is being reflected from the truncated disk.  As the {\ttfamily diskir} component
must produce the high energy emission in this case, a very high Comptonization 
temperature is required ($kT_{e} > 429$\,keV) and the Comptonizing fraction 
increases to $L_{c}/L_{d} = 2.4\pm 0.6$.  The {\ttfamily reflionx} parameters are 
similar to those described above for the X-ray only fits.  The ionization state 
is low, with a value of $\xi = 5.0^{+4.4}_{-2.2}$ erg\,cm\,s$^{-1}$.  Also, the
Fe abundance is $0.33\pm 0.09$, and the covering fraction is $\Omega/2\pi = 0.20$.
The ionization parameter and the covering fraction do not seem unreasonable for a 
cool and truncated disk, but we cannot say with any certainty which two-component 
model is more likely to be correct.

\subsubsection{Full SED}

When the full radio to hard X-ray SED is put together, it is immediately clear that 
the extrapolation of the power-law seen in the radio band is well above the flux 
measured in the near-IR (even after dereddening).  This implies that the radio 
component, which is attributed to the compact jet, must have a spectral break between
the IR and radio bands.  Thus, we fit the SED with a model consisting of a
broken power-law ({\ttfamily bknpower}) and a {\ttfamily diskir} component.  The 
{\ttfamily bknpower} component provides all of the emission in the radio band, and we 
fix the power-law index below the break energy ($E_{\rm break}$) to $\Gamma_{1} = 0.7$.  
The index above the break ($\Gamma_{2}$) is left as a free parameter, and we find that 
the best fit model has a strong contribution from the {\ttfamily bknpower} above 
$\approx$20\,keV.  As described above, the GSO data require a cutoff, and we multiplied 
the broken power-law component with a high-energy cutoff.  

The continuum components are multiplied by {\ttfamily redden} and {\ttfamily tbabs} as 
described above.  We fixed $E(B-V)$ to 0.45 and $N_{\rm H}$ was left as a free parameter.  
The fit parameters are given in Table~\ref{tab:parameters_sed}, and the quality of the 
fit is $\chi^{2}_{\nu} = 1.28$ for 2156 dof.  We left the normalizations between the 
X-ray instruments as free parameters, but we fixed all of the non-X-ray instrument 
normalizations to the {\em Swift}/XRT normalization.  

\begin{figure}
\hspace{-0.2cm}
\includegraphics[scale=0.5]{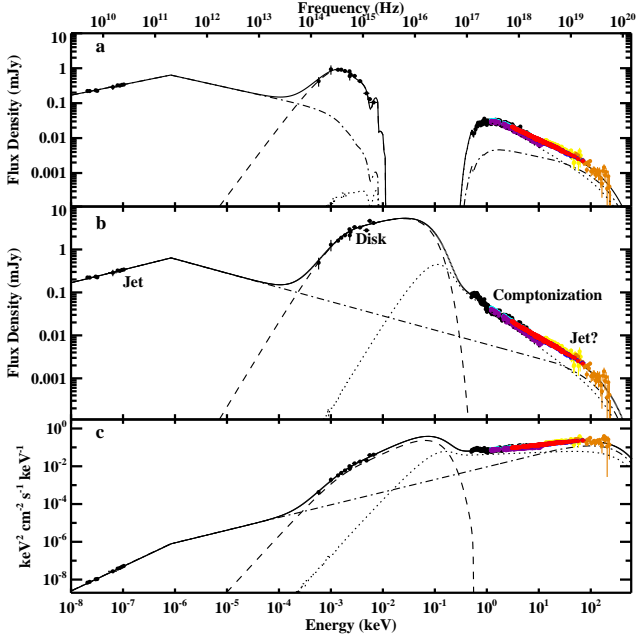}
\vspace{0.2cm}
\caption{\small {\it (a)} The radio to hard X-ray spectral energy distribution 
for Swift~J1753.5--0127.  The model is a broken power-law with a high-energy 
cutoff (dash-dotted line) and a {\ttfamily diskir} component, which we have
divided into its thermal component (dashed line) and its Comptonization 
component (dotted line).  The points are not dereddened, and we use
$E(B-V) = 0.45$ and $N_{\rm H} = 2.84\times 10^{21}$\,cm$^{-2}$ for the model.
The solid line is the sum of the components. {\it (b)} The same data and model 
after dereddening. {\it (c)} The same data and model multiplied by energy.
\label{fig:spectrum_all}}
\end{figure}

We used the XSPEC routine {\ttfamily steppar} to determine the range of possible values 
for $E_{\rm break}$.  The $\chi^{2}$ values are nearly constant over a large range, increasing 
sharply at $1.0\times 10^{-7}$\,keV ($2.4\times 10^{10}$\,Hz), which corresponds to the 
highest radio frequency measured, and at $1.5\times 10^{-5}$\,keV ($3.6\times 10^{12}$\,Hz).  
At the upper limit, $\Gamma_{2}$ becomes steeper to avoid over-producing in the near-IR, but
$\chi^{2}$ becomes worse because the component no longer extends to the X-ray band.  
For Figure~\ref{fig:spectrum_all}, showing the fitted SED, we set $E_{\rm break}$ to 
$1.0\times 10^{-6}$\,keV as an example.  The main result is that it is possible for the
broken power-law to account for the hard X-ray excess.  The {\ttfamily diskir} parameters
for the full SED fit (see Table~\ref{tab:parameters_sed}) are almost the same as 
the parameters for the {\ttfamily diskir+highecut*pegpwrlw} fits to the near-IR to X-ray
fits (see Table~\ref{tab:parameters2}).

If the hard X-ray excess is explained by a reflection component instead of the broken
power-law (i.e., the jet), then $\Gamma_{2}$ could be steeper, allowing for even higher
values of $E_{\rm break}$.  We explored this possibility by fitting just the radio to UV
spectrum with a modified version of the model shown in Table~\ref{tab:parameters_sed}.
The modifications include removing {\ttfamily highecut} and fixing the {\ttfamily diskir}
components related to Comptonization to the values found for the full SED.  In addition, 
while we allowed $\Gamma_{2}$ to be a free parameter, we did not allow this part of the
broken power-law to be steeper than $\Gamma_{2} = 2$.  While the lower limit on $E_{\rm break}$
is unchanged, the upper limit moves higher, and values as high as 
$E_{\rm break} = 6.5\times 10^{-5}$\,keV ($1.6\times 10^{13}$\,Hz) are possible.

\subsection{X-ray Timing}

\begin{figure}
\hspace{-0.2cm}
\includegraphics[scale=0.52]{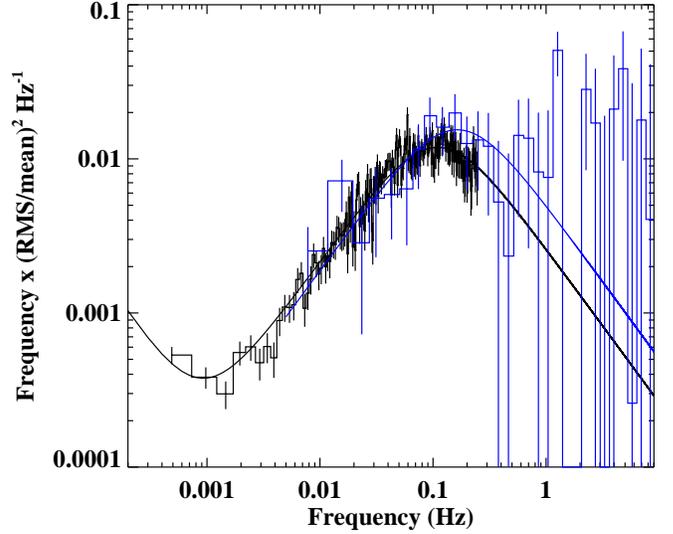}
\caption{\small Soft X-ray power spectrum from {\em Suzaku}/XIS (black) and 
{\em Swift}/XRT (blue) fitted with a zero-centered Lorentzian and a low-frequency 
power-law.
\label{fig:power}}
\end{figure}

We made power spectra using the {\em Suzaku}/XIS and {\em Swift}/XRT data.  XIS has a larger 
effective area and the exposure time is much longer than XRT, so the statistical quality is 
much better.  However, the XRT data are useful because of the higher time resolution.  There 
is good agreement between the two power spectra in the frequency region where they overlap 
(see Figure~\ref{fig:power}), but precise agreement is not expected due to the different times 
being covered and the slightly different energy bandpasses.  Thus, we fitted the power spectra 
separately.  For XIS, we used a zero-centered Lorentzian and a power-law at low frequencies.  
For XRT, the zero-centered Lorentzian is sufficient.  The parameters are shown in 
Table~\ref{tab:parameters_power}, and the fractional rms of the Lorentzians are 27.3\%$\pm 0.2$\% 
for XIS and 22\%$\pm 2$\% for XRT, which is consistent with the relatively high levels of 
variability expected for the hard state.  The full-width at half-maximum (FWHM) of the Lorentzians 
are $0.220\pm 0.005$\,Hz for XIS and $0.33\pm 0.07$\,Hz for XRT.  In previous work on timing 
analysis of Swift~J1753.5--0127 \citep{soleri13,kalamkar15}, the Lorentzian fits were characterized 
by the frequency where the power spectrum is maximal when plotted as frequency times rms power 
($\nu_{\rm max}$) as shown in Figure~\ref{fig:power}.  For XIS and XRT, the values of $\nu_{\rm max}$
are $0.110\pm 0.003$\,Hz and $0.16\pm 0.04$\,Hz, respectively.  

\section{Discussion}

In this work, we have performed detailed spectral fits to the most complete SED that has been 
obtained for Swift~J1753.5--0127 to date.  While previous multi-wavelength studies of this
source that included radio measurements have covered the radio, near-IR, optical, and X-ray 
\citep{cadollebel07,durant09,reynolds10,soleri10,zyc10}, we have obtained radio detections at nine 
frequencies, included UV coverage, and used a combination of seven X-ray spectra, covering 
0.5\,keV to 240\,keV.  Here, we discuss three main topics:  1. the implications of the constraint 
on $\nu_{\rm break}$ for the compact jet properties; 2. what we can infer about the properties of 
the optically thick accretion disk; and 3. the possible origins of the high-energy emission 
components.

For all these topics, it is useful to estimate the luminosity of Swift~J1753.5--0127 during 
these observations.  For the model shown in Figure~\ref{fig:spectrum_all}, the absorbed flux
over the full energy band covered ($2\times 10^{-8}$\,keV to 240\,keV) is 
$1.25\times 10^{-9}$\,erg\,cm$^{-2}$\,s$^{-1}$.  Although there is uncertainty about the
break frequency of the broken power-law, this leads to very little uncertainty in the flux
since essentially all of the flux is above 1\,eV.  The unabsorbed flux is 
$2.71\times 10^{-9}$\,erg\,cm$^{-2}$\,s$^{-1}$ in the 1\,eV to 240\,keV band, and this 
represents the bolometric flux.  This is for the model in Table~\ref{tab:parameters_sed}, 
but the unabsorbed flux for the {\ttfamily diskir+reflionx} model shown in 
Table~\ref{tab:parameters2} gives an unabsorbed flux of $2.38\times 10^{-9}$\,erg\,cm$^{-2}$\,s$^{-1}$
due to the lower column density.  Using the average of these two unabsorbed fluxes, 
the bolometric luminosity is $2.7\times 10^{36}$~$d_{3}^{2}$\,erg\,s$^{-1}$, where 
$d_{3}$ is the distance to the source in units of 3\,kpc.  For a black hole mass of 
5\Msun, this corresponds to an Eddington-scaled luminosity of 0.41\%~$d_{3}^{2}$~$M_{5}^{-1}$, 
where $M_{5}$ is the black hole mass in units of 5\Msun.

\subsection{The Compact Jet and the Break Frequency}

We are able to obtain a constraint on $\nu_{\rm break}$ because of the rising and well 
constrained radio spectrum ($\alpha = 0.29\pm 0.05$) along with the fact that the spectrum 
rises from $K_{s}$-band to higher frequencies.  Without considering the X-rays, we find 
that $\nu_{\rm break}<1.6\times 10^{13}$\,Hz ($\log{\nu_{\rm break,Hz}} < 13.2$).  If the jet 
does contribute to the X-rays, then $\nu_{\rm break}<3.6\times 10^{12}$\,Hz 
($\log{\nu_{\rm break,Hz}} < 12.6$).  A study of 16 $\nu_{\rm break}$ measurements or limits 
for nine black hole systems in the hard state found mostly higher values than the 
Swift~J1753.5--0127 upper limits \citep{russell13a}.  For the measurements, the median 
value of $\log{\nu_{\rm break,Hz}}$ is 13.68, and the values range from 12.65 (for XTE~J1118+480) 
to 14.26 (for GX~339--4 and V404~Cyg).  When limits are also considered, there are still 
only two measurements that are as low as the value found for Swift~J1753.5--0127:  
$\log{\nu_{\rm break,Hz}} = 12.65\pm 0.08$ for XTE~J1118+480 and $<$13.13 for GX~339--4.  

While relatively low, the single $\nu_{\rm break}$ measurement for Swift~J1753.5--0127 
does not necessarily indicate anything unusual about the system itself.  Multiple
measurements of individual systems show significant changes for GX~339--4, XTE~J1118+480, 
MAXI~J1836--194, and MAXI~J1659--152 \citep{gandhi11,russell13a,russell13b,vanderhorst13}.  
For GX~339--4, \cite{gandhi11} found that $\nu_{\rm break}$ changed by a factor of 
$>$10 in less than a day.  For MAXI~J1836--194, six measurements over a period of less 
than two months showed changes in $\log{\nu_{\rm break,Hz}}$ from close to 11 to close to 
14 while the source changed X-ray luminosity and hardness \citep{russell13b,russell14}, 
and the highest value of $\nu_{\rm break}$ occurred when the source was at its lowest X-ray 
luminosity with its hardest X-ray spectrum.  The Swift~J1753.5--0127 measurements occurred 
when the spectrum was hard and the X-ray luminosity was low; thus, it may not follow the 
same trend as MAXI~J1836--194.  However, this is not surprising since the larger source 
sample studied in \cite{russell13a} did not show any evidence for a correlation between 
X-ray luminosity and $\nu_{\rm break}$.

In the canonical model for compact jets \citep{bk79}, the jet spectrum is composed of
a superposition of synchrotron components with a continuum of peak frequencies due to
changing optical depth.  The synchrotron spectrum from each region depends primarily
on the magnetic field strength and also on the radial size of the jet.  The value of 
$\nu_{\rm break}$ depends on both the magnetic field and the radial size of the jet
in its acceleration zone, which is close to the base of the jet.  To place constraints
on these quantities ($B$ and $R$), we use equations 1 and 2 from \cite{gandhi11}, which
are based on a single-zone cylindrical approximation \citep{cdr11}.  We estimate 
the upper limit on $B$ using the parameters from the full SED fit (see 
Table~\ref{tab:parameters_sed}).  The input parameters to the equations are
$\nu_{\rm break}<3.6\times 10^{12}$\,Hz, the flux at $3.6\times 10^{12}$\,Hz, which is 
1.42\,mJy, and the slope of the power-law above $\nu_{\rm break}$.  To determine the
slope, we fixed $\nu_{\rm break}$ to $3.6\times 10^{12}$\,Hz, refit the SED, and found
a value of 1.4, which corresponds to $\alpha$ = --0.4.  The upper limit on the 
magnetic field strength in the acceleration zone is $B < 2.4\times 10^{3}$~$d_{3}^{-0.24}$\,G
and $R > 1.8\times 10^{9}$~$d_{3}^{0.936}$\,cm.  If we do not consider the X-rays, 
$\nu_{\rm break}<1.6\times 10^{13}$\,Hz, the flux at $1.6\times 10^{13}$\,Hz is 2.18\,mJy, 
and the slope of the power-law above $\nu_{\rm break}$ is assumed to be 2 ($\alpha$ = --1), 
giving $B < 9.6\times 10^{3}$~$d_{3}^{-0.21}$\,G and $R > 4.6\times 10^{8}$~$d_{3}^{0.954}$\,cm.
Also, from the radio alone, we know that $\nu_{\rm break}>2.5\times 10^{10}$\,Hz, and the
flux at this frequency is 0.34\,mJy.  Assuming $\alpha$ = --1, we derive 
$B > 18$~$d_{3}^{-0.21}$\,G and $R < 1.2\times 10^{11}$~$d_{3}^{0.954}$\,cm.

Two examples of hard state black hole systems for which $B$ and $R$ have been previously 
calculated using this same technique are GX~339--4 \citep{gandhi11} and MAXI~J1836--194 
\citep{russell14}.  For GX~339--4, these quantities were estimated to be $B\approx 1.5\times 10^{4}$\,G
and $R\approx 2.5\times 10^{9}$\,cm.  For MAXI~J1836--194, estimates for $B$ and $R$ were 
obtained for three hard state observations: one during the rise of an outburst and two during 
outburst decay.  Figure~6 of \cite{russell14} shows $B\sim 10^{2}$\,G and $R\sim 10^{12}$\,cm 
during the rise and $B\sim 3\times 10^{3-4}$\,G and $R\sim 10^{9-10}$\,cm during the decay.  
Thus, the ranges of $B = 1.8\times 10^{1}$--$9.6\times 10^{3}$~$d_{3}^{-0.21}$\,G and 
$R = 4.6\times 10^{8}$--$1.2\times 10^{11}$~$d_{3}^{0.954}$\,cm that we derive for 
Swift~J1753.5--0127 are largely consistent with the range of values previously determined
for these two sources.  The best agreement in the jet properties between Swift~J1753.5--0127
and GX~339--4 and MAXI~J1836--194 (during decay) occurs if the actual value of $\nu_{\rm break}$ 
for Swift~J1753.5--0127 is close to the upper end of the range of possible values.

\subsection{The Optically Thick Accretion Disk}

Here, we discuss the spectral components that can be modeled as thermal emission and
the implications for the optically thick accretion disk.  First, we discuss the
near-IR to UV component that is consistent with a multi-temperature disk model with 
$kT_{\rm in} = 28^{+21}_{-11}$\,eV.  Then, we consider the possibility of a second 
thermal component in the soft X-ray band with $kT_{\rm in}\approx$ 150\,eV.  

Our spectral model assumes that the near-IR to UV emission is strongly dominated
by a disk component, and it is worthwhile to consider how secure this assumption 
is.  We know that at least a large fraction of the emission comes from the disk 
because of the double-peaked emission lines that are seen in this bandpass
(Froning et al. 2014\nocite{froning14}; Neustroev et al. 2014\nocite{neustroev14}; 
Rahoui et al., submitted to ApJ).  However, \cite{neustroev14} also find
a weak emission line and two weak absorption lines (all three unidentified) in the 
optical, which they interpret as coming from the companion star.  If this interpretation 
is correct (and we note that the fiducial black hole mass and source distance that we 
use in this paper depend on it), then it requires some contribution from the companion 
in the optical.  Without X-ray irradiation, the emission from the companion would 
be negligible: a blackbody with a temperature of 3000\,K \citep{neustroev14}, a 
radius equal to the companion's Roche lobe size of $1.68\times 10^{10}$\,cm, and a 
distance of 3\,kpc has a flux that is two orders of magnitude lower than the 
measured flux in the near-IR and three orders of magnitude lower in the optical.  
Thus, the temperature of the irradiated side of the companion must be significantly
hotter for there to be a contribution to the optical flux.  However, the crucial
point is that even if the three lines are from the companion, they are extremely
weak in comparison to the very strong double-peaked lines from the disk, indicating
that the disk emission is much stronger than any potential contribution from the
companion.

One possibility that we cannot completely rule out is that there are additional 
components from the compact jet.  The broken power-law emission represents the 
post-shock synchrotron component.  While this is the only component that has been 
seen in SEDs of accreting black holes that is widely accepted as emission from the 
compact jet, theoretical jet models indicate that pre-shock synchrotron can be 
relatively bright in the optical and UV \citep{mnw05,homan05,migliari07,maitra09}.
Another possibility that has been suggested as a contributor to the optical emission
is synchrotron emission from non-thermal electrons in the hot accretion flow (i.e., 
the corona).  A complex optical/X-ray cross-correlation function was reported for
Swift~J1753.5--0127 \citep{durant08,durant11}, and it was shown that it could be
explained if the optical emission had components from the disk and the corona
\citep{vpv11}.  The coronal contribution to the cross-correlation function has 
been observed to vary inversely with the strength of the disk \citep{hynes09a}.
Rahoui et al. (submitted to ApJ) show that Swift~J1753.5--0127 had a strong and 
likely dominant thermal disk component in observations taken a few months after 
ours\footnote{The Rahoui et al. (submitted to ApJ) observations were made on 
2014 August 16 (MJD 56885), and the X-ray light curves shown in Figure~\ref{fig:lc} 
do not show any major change between April and August.}, which would suggest a 
relatively weak coronal contribution to the optical during our observations.

With the caveats about the possibility of a fractional contribution from pre-shock 
synchrotron emission or the corona, we can compare the parameters of the thermal 
near-IR to UV component to previous studies of Swift~J1753.5--0127 SEDs where this 
component has also been modeled as thermal emission \citep{zyc10,froning14}.  
From Tables~\ref{tab:parameters2} (last two columns) and \ref{tab:parameters_sed}, 
the values of $N_{\rm diskbb}$ are $(5.5^{+18}_{-3.1})\times 10^{7}$, 
$(8.2^{+72}_{-6.1})\times 10^{7}$, and $(9.0^{+89}_{-1.2})\times 10^{7}$.  Using
Equation~1, these values imply strongly truncated disks.  As before, we
assume $M_{\rm BH} = 5$\Msun, $d = 3$\,kpc, and $i = 40^{\circ}$.  To determine
the lower limits on the inner disk radii, we assume $f = 1$, and the values
are $R_{\rm in} > 227$\,$R_{g}$, $> 212$\,$R_{g}$, and $> 409$\,$R_{g}$.
While previous studies have mostly assumed a larger distance to 
Swift~J1753.5--0127, this would make the values of $R_{\rm in}$ larger.
\cite{froning14} modeled a near-IR to UV SED and determined that $R_{\rm in}$ 
needed to be $> 100$\,$R_{g}$ to avoid overpredicting the simultaneously 
measured X-ray spectrum.  \cite{zyc10} used a self-consistent model with
optically thick disk emission, jet emission, and a Comptonization component,
and they were able to fit a radio to hard X-ray SED with $R_{\rm in} = 500$\,$R_{g}$.
\cite{zyc10} assumed different values for $d$, $M_{\rm BH}$, and $i$, and if
we recalculate their $R_{\rm in}$ using the values we adopt, the result
is $R_{\rm in} = 350$\,$R_{g}$.  While the precise value of $R_{\rm in}$ is 
likely to vary in time, all of these measurements suggest that the near-IR 
to UV component comes from a strongly truncated disk.

The spectral fits also constrain the outer disk radius based on the parameter
$\log{(R_{\rm out}/R_{\rm in})} = 2.31^{+0.06}_{-0.04}$ (see Table~\ref{tab:parameters_sed}).
For $N_{\rm diskbb} = 9\times 10^{7}$ (the best fit value), we calculate 
$R_{\rm out} = 6.6\times 10^{10}$\,cm.  We compare this value to the system
parameters reported by \cite{neustroev14}, where they determine that the 
binary separation is $a\lsim 1.1\times 10^{11}$\,cm, and the size of the 
black hole's Roche lobe is $7.1\times 10^{10}$\,cm.  A filling fraction of 
90\% is typically assumed for an accretion disk, which would result in a
predicted disk size of $6.4\times 10^{10}$\,cm, which is in excellent agreement
with our measurement.  Although it will be important to confirm the system
parameters with radial velocity measurements of the companion star when
the source is in quiescence (if it is bright enough), we see this $R_{\rm out}$
comparison as another piece of evidence that the near-IR to UV component is
strongly dominated by emission from the accretion disk.

The 150\,eV component is marginally significant in the XRT spectrum, and 
it is not detected when the XIS data are included.  However, for previous
observations of Swift~J1753.5--0127, the presence of a 0.1--0.4\,keV thermal 
component was well-established from spectral (see references in Section 3.1.3)
and timing \citep{uttley11} measurements.  Even though our 2014 April observation
is at a moderately lower X-ray flux level (only a factor of 2--3 lower than
the majority of the previous observations), seeing a weak thermal component
in the X-ray band is not surprising.  If we use $N_{\rm diskbb} = 1.1\times 10^{4}$ 
and carry out the same inner radius calculation as performed for the near-IR 
to UV component, we obtain $R_{\rm in} = 5$\,$R_{g}$ for $f = 1$ and 
$R_{\rm in} = 14$\,$R_{g}$ for $f = 1.7$, suggesting that this component could
come from a disk that extends close to the ISCO.  The presence of two 
thermal components in the SED of Swift~J1753.5--0127 has been previously
reported \citep{chiang10}, and potential physical interpretations are
discussed in that work.  It has been shown that a small inner optically
thick accretion disk can form due to condensation of material from the
corona \citep{liu07,mlm07,taam08}, and \cite{chiang10} consider this
possibility as well as a scenario where strong irradiation at the inner
edge of a truncated disk distorts the temperature profile.  For the inner
disk possibility, it has been predicted that the inner disk can exist
down to $L/L_{\rm Edd}\sim 0.1$\% and then completely evaporate below
this level \citep{taam08}.  Thus, given the luminosity of Swift~J1753.5--0127
during our observation ($L/L_{\rm Edd}\sim 0.4$\%), the presence of an
inner disk is predicted.

As previously mentioned, the luminosity at the time of our observation of 
Swift~J1753.5--0127 was close to the lowest level since the source was
discovered, but it was only a factor of a few times lower than the highest
levels seen over the past several years (see Figure~\ref{fig:lc}).  
The X-ray power spectrum also suggests that the properties during our
observation were at one end of a continuum as opposed to requiring some
major overall change in the system.  \cite{soleri13} report on timing
analysis of 67 {\em Rossi X-ray Timing Explorer (RXTE)} observations 
of Swift~J1753.5--0127 during 2009 and 2010.  While the comparison to
our observation is somewhat complicated by the fact that most of the
{\em RXTE} observations required two Lorentzian components, fifteen
of the power spectra were fitted with a single Lorentzian, allowing
for a direct comparison.  For those cases, the values of $\nu_{\rm max}$
range from 0.18\,Hz to 3.18\,Hz.  Thus, our {\em Suzaku} and {\em Swift}
measurements of $0.110\pm 0.003$\,Hz and $0.16\pm 0.04$\,Hz, respectively,
are only slightly lower than the \cite{soleri13} measurements.  

\subsection{The Origin of the X-ray Emission}

A major question in recent years concerns how much of the X-ray emission
can be attributed to the compact jet.  In the model of \cite{mnw05}, the
jet can produce X-rays via post-shock synchrotron emission, which can be
modeled as the broken power-law that we use in our fits, or synchrotron
self-Compton (SSC) from the base of the jet, which can contribute in the
hard X-ray band.  The SEDs of GX~339--4, GRO~J1655--40, and XTE~J1118+480
allow for the possibility that all the soft X-ray emission comes from 
the post-shock synchrotron component \citep{mnw05,migliari07,maitra09}.
For Swift~J1753.5--0127, Figure~\ref{fig:spectrum_all} shows that such a
scenario is ruled out, and a Comptonization component is strongly required
by the data. 

The question of what makes Swift~J1753.5--0127 different is directly 
relevant to the question of what is different about the outliers in the 
X-ray/radio correlation.  Although one possibility is that Swift~J1753.5--0127
has a stronger Comptonization component in the X-rays, another possibility
is that it has a weaker radio jet.  In Section 4.1, we showed that the 
highest possible peak flux for the Swift~J1753.5--0127 broken power-law
component is 2.18\,mJy.  This corresponds to a specific (peak) luminosity 
of $2.3\times 10^{19}$$d_{3}^{2}$\,erg\,s$^{-1}$\,Hz$^{-1}$ at
$\nu_{\rm break} = 1.6\times 10^{13}$\,Hz.  \cite{russell13a} give 15
peak luminosities for nine hard state black hole systems, and the values
range from $7.1\times 10^{19}$ to $1.9\times 10^{22}$\,erg\,s$^{-1}$\,Hz$^{-1}$
with a median value of $1.2\times 10^{21}$\,erg\,s$^{-1}$\,Hz$^{-1}$.  Thus, 
assuming a distance of 3\,kpc, the peak jet luminosity for Swift~J1753.5--0127 
is 50 times lower than the median and 3 times lower than the least luminous
system.  The distance to Swift~J1753.5--0127 would need to be 5--6\,kpc to 
to move the Swift~J1753.5--0127 peak jet luminosity close to the least luminous 
system, which is conceivable, but it would need to be $\approx$21\,kpc to make 
the Swift~J1753.5--0127 comparable with the median peak jet luminosity, which 
can be ruled out.

We made a second radio luminosity comparison by integrating the radio power-law
measurements for Swift~J1753.5--0127 and the black hole sources from
\cite{russell13a} up to $\nu_{\rm break}$.  For Swift~J1753.5--0127, the 
luminosity up to $1.6\times 10^{13}$\,Hz is $3.1\times 10^{32}$$d_{3}^{2}$\,erg\,s$^{-1}$.  
For the sources from \cite{russell13a}, not all 15 of the SEDs are high enough
quality to make a reliable luminosity determination.  There was sufficient information
to calculate ten luminosities for eight sources.  These ranged from 
$1.8\times 10^{33}$\,erg\,s$^{-1}$ for XTE~J1118+480 and $2.0\times 10^{33}$\,erg\,s$^{-1}$
for Cyg~X-1 to $1.1\times 10^{36}$\,erg\,s$^{-1}$ for GS~1354--64 and 
$3.1\times 10^{36}$\,erg\,s$^{-1}$ for V404~Cyg.  The median value is 
$1.1\times 10^{35}$\,erg\,s$^{-1}$, and the distance to Swift~J1753.5--0127 is
certainly not large enough for the luminosity to approach that value.  Thus, the 
low luminosity radio jet may be at least part of the reason why Swift~J1753.5--0127 
is an outlier.

While the Swift~J1753.5--0127 SED is consistent with Comptonization being
dominant at soft X-rays, our results show that multiple components are 
required to explain the entire 0.5--240\,keV X-ray spectrum.  In our 
spectral fits, we considered a reflection component or the post-shock
synchrotron component.  Figure~\ref{fig:nir2x2} illustrates the reflection
possibility, and such a scenario is consistent with our overall picture for
the system.  The outer optically thick disk could produce a weak 
($\Omega/2\pi = 0.20$) reflection component, and it would be expected to have 
a low ionization, which is consistent with $\xi = 5.0^{+4.4}_{-2.2}$\,erg\,cm\,s$^{-1}$.
While an iron line detection would be strong evidence in favor of the reflection
interpretation, there is no iron line in the Swift~J1753.5--0127 spectrum, but
we find that it is possible to explain the lack of an iron line with an iron 
abundance of $0.33\pm 0.09$ of the solar value.  This iron abundance may
be problematic for the reflection interpretation, but we do not think that
it is low enough to rule it out.  We have also considered the fact that this 
is the only model that requires a very high Comptonization temperature 
($kT_{\rm e} > 429$\,keV).  This occurs because the reflection component falls 
at high energies, allowing the overall model to fit the steeper {\em Suzaku}/GSO 
spectrum without an exponential cutoff in the direct model.  This electron 
temperature is higher than has been inferred from measurements of other accreting 
black holes in the higher luminosity parts of their hard states, which are 
typically in the 50--120\,keV range \citep[][and references therein]{pv14}.  
However, it is predicted that $kT_{\rm e}$ should increase to hundreds of keV in 
the lower luminosity parts of the hard state \citep{gd13}.  Thus, the lack of an 
iron line is the strongest reason to disfavor the reflection possibility, but 
this scenario is not ruled out.

The model where the hard X-rays are due to post-shock synchrotron emission
(see Figure~\ref{fig:nir2x1}) has the advantage of a much more typical electron 
temperature ($kT_{\rm e} > 33$\,keV).  On the other hand, the slope of the 
power-law, $\Gamma = 1.33^{+0.08}_{-0.25}$ ($\alpha$ = --$0.33^{+0.25}_{-0.08}$),
while not unreasonable for optically thin synchrotron emission, is 
harder than is seen for other black hole systems, which have values of
$\alpha$ between --0.68 and --1.38 \citep{russell13a}.  Such a hard spectrum
also requires that the spectrum is sharply cutoff to explain the steeper
{\em Suzaku}/GSO spectrum, and the exponential cutoff with $E_{\rm cut} = 20\pm 3$\,keV
and $E_{\rm fold} = 142^{+110}_{-25}$\,keV is probably inconsistent with the
more gradual cutoff predicted for a synchrotron spectrum \citep{zdziarski03}.
A third possibility that was mentioned above but not specifically considered in
our spectral modeling is that the extra hard X-ray component is due to SSC
emission from the base of the jet.  Fits with the \cite{mnw05} compact jet
model are beyond the scope of this paper, but it would be interesting to 
use our SED to test this model in future work.

Finally, we have considered whether any of our conclusions might be affected 
by day-to-day source variability given that the observations we use for the full 
SED cover $\sim$2.8 days from the $K_{s}$-band observation to the VLA observation
(although most of the measurements for the SED come from a smaller span of times).
Figure~\ref{fig:lc_zoom} shows that there is little day-to-day variability in the
optical and near-IR during the campaign, and this is consistent with previous
long-term studies of Swift~J1753.5--0127 \citep[e.g.,][]{shaw13}.  Also, the
{\em Suzaku}/XIS observations show day-to-day stability in the soft X-ray flux.
It is a little less clear whether there are changes in the radio and IR compact
jet spectrum as other black hole systems have shown significant changes in the 
break frequency on time scales of a day as discussed in Section 4.1.  We already
consider a large range of break frequencies; thus, the conclusions that there
is a separate thermal component in the near-IR to UV and that a Comptonization
component is required in the soft X-ray should not be affected.  However, the
question of whether the extra hard X-ray component comes from the compact jet
depends very sensitively on the break frequency and spectral slope.  To reach
a definitive conclusion on the origin of the hard X-ray emission may require
simultaneous radio and hard X-ray monitoring.

\section{Summary and Conclusions}

We have obtained radio, near-IR, optical, UV, and X-ray coverage for the long-term
black hole transient Swift~J1753.5--0127 in 2014 April when the source was in the
hard state at one of its lowest X-ray luminosities ($2.7\times 10^{36}$~$d_{3}^{2}$\,erg\,s$^{-1}$)
since the discovery of the source.  We performed fitting of the broadband energy 
spectrum as well as the X-ray power spectrum.  We obtain results concerning the
compact jet, the optically thick accretion disk, and the origin of the X-ray 
emission, which is also relevant for the question of why Swift~J1753.5--0127 is
a radio/X-ray correlation outlier.

With the combination of the rising radio spectrum, and the rise in the near-IR, 
$\nu_{\rm break}$ is constrained for the post-shock synchrotron component of the
compact jet, and this provides constraints on $B$ and $R$ for the jet acceleration 
zone.  While the post-shock synchrotron component may contribute in hard X-rays, 
the soft X-ray flux is far too high to be part of this component, which we model
with a Comptonization component.  Based on this result, Swift~J1753.5--0127
appears to be an outlier because of the combination of a strong Comptonization
component and a jet with peak and broadband luminosities significantly lower
than is seen for other black hole systems.

The low jet luminosity and the low extinction for Swift~J1753.5--0127 appear to 
provide an opportunity to clearly see emission components that may be too weak
or too absorbed to see in other systems.  The double-peaked emission lines
(Froning et al. 2014\nocite{froning14}; Neustroev et al. 2014\nocite{neustroev14}; 
Rahoui et al., submitted to ApJ) clearly show that the near-IR to UV spectrum has at
least a strong (likely dominant) thermal disk component.  Further evidence that
the near-IR is dominated by thermal disk emission is that the component can be
modeled by a disk with an outer radius of $R_{\rm out}/R_{g} = 90,000$\,$d_{3}$\,$M_{5}^{-1}$
($R_{\rm out}=6.6\times 10^{10}$\,$d_{3}$\,cm), consistent with the expected size of 
the disk given previous measurements of the size of the companion's Roche lobe.  
The fact that this component does not contribute in the X-ray band constrains the 
inner radius to be $R_{\rm in}/R_{g} > 212$\,$d_{3}$\,$M_{5}^{-1}$.  While this implies 
that the near-IR to UV emission comes from a strongly truncated disk, there is also 
some evidence for a weak 150\,eV thermal component in the soft X-rays, and its inner 
radius could be as small as 5--14\,$R_{g}$.  The presence of two thermal components 
could provide support for predictions that low luminosity systems may have inner
and outer optically thick disks with a gap in the middle.

Finally, we have considered the possibility that there is a reflection 
component in the spectrum.  In the presence of strong hard X-rays, one expects 
to see a reflection component from the optically thick material.  The hard X-ray 
spectrum is consistent with the presence of a reflection component, but no iron 
line is detected.  The low ionization ($\xi = 5.0^{+4.4}_{-2.2}$\,erg\,cm$^{-2}$\,s$^{-1}$)
and low covering fraction ($\Omega/2\pi = 0.2$) would favor the possibility that
this component comes from the outer optically thick disk.  If reflection is the
cause of the second hard X-ray component, then invoking the jet to explain the
extra hard X-ray emission (see Figure~\ref{fig:spectrum_all}) may not be 
required.

\acknowledgments

We thank the referee for useful comments that helped to improve the manuscript.  
This work was supported under NASA Contract No. NNG08FD60C, and made use of data 
from the {\it NuSTAR} mission, a project led by  the California Institute of 
Technology, managed by the Jet Propulsion  Laboratory, and funded by the National 
Aeronautics and Space Administration. We thank the {\it NuSTAR} Operations, 
Software and  Calibration teams for support with the execution and analysis of 
these observations.  This research has made use of the {\it NuSTAR} Data Analysis 
Software (NuSTARDAS) jointly developed by the ASI Science Data Center (ASDC, Italy) 
and the California Institute of  Technology (USA).  The PS1 Surveys have been made 
possible through contributions of the Institute for Astronomy, the University of 
Hawaii, the Pan-STARRS Project Office, the Max-Planck Society and its participating
institutes, the Max Planck Institute for Astronomy, Heidelberg and the Max Planck 
Institute for Extraterrestrial Physics, Garching, The Johns Hopkins University, 
Durham University, the University of Edinburgh, Queen's University Belfast, the 
Harvard-Smithsonian Center for Astrophysics, and the Las Cumbres Observatory Global 
Telescope Network, Incorporated, the National Central University of Taiwan, and the 
National Aeronautics and Space Administration under Grant No. NNX08AR22G issued 
through the Planetary Science Division of the NASA Science Mission Directorate.  
JAT acknowledges partial support from NASA under {\em Swift} Guest Observer grants 
NNX13AJ81G and NNX14AC56G.  SC acknowledges the financial support from the UnivEarthS 
Labex programme of Sorbonne Paris Cit\'e (ANR-10-LABX-0023 and ANR-11-IDEX-0005-02), 
and from the CHAOS project ANR-12-BS05-0009 supported by the French Research National 
Agency.  JMJ is supported by an Australian Research Council (ARC) Future Fellowship 
(FT140101082) and also acknowledges support from an ARC Discovery Grant.  EK 
acknowledges support from TUBITAK BIDEB 2219 program.  This work was supported by 
the Spanish Ministerio de Econom\'a y Competitividad (MINECO) under grant 
AYA2013-47447-C3-1-P (SM).



\begin{table}
\caption{Observing Log and Exposure Times\label{tab:obs}}
\begin{minipage}{\linewidth}
\begin{center}
\footnotesize
\begin{tabular}{ccccccc} \hline \hline
        &            &   &  & Start Time (UT) & End Time (UT) & Exposure\\
Mission & Instrument & Energy/Filter & ObsID & (in 2014)       & (in 2014)     & (s)\\
\hline\hline
\multicolumn{7}{c}{Radio}\\ \hline
VLA & --- & 4--8\,GHz      & --- & Apr 5, 11.00 h      & Apr 5, 13.00 h & 1518\\
VLA & --- & 18--26\,GHz    & --- &     ''              &       ''       & 1746\\
AMI  & --- & 12--17.9\,GHz & --- & Apr 4, 3.07 h       & Apr 4, 7.56 h  & 16,164\\
AMI  & --- & 12--17.9\,GHz & --- & Apr 5, 2.67 h       & Apr 5, 7.64 h  & 17,892\\ \hline
\multicolumn{7}{c}{Near-IR to UV}\\ \hline
Kanata & HONIR & $B$/$V$/$J$/$K_{s}$ & --- & Apr 2, 17.3 h & Apr 7, 19.2 h & see text and Table~\ref{tab:obs_kanata}\\
SLT & U42 & $g^{\prime}$/$r^{\prime}$/$i^{\prime}$/$z^{\prime}$ & --- & Apr 3, 9.1 h & Apr 5, 11.0 h & 180\footnote{Each exposure listed in Table~\ref{tab:obs_slt} was 180\,s.}\\
{\em Swift} & UVOT & $v$    & 00080730001 & Apr 5, 0.4 h & Apr 5, 5.4 h & 182\\
{\em Swift} & UVOT & $b$    & 00080730001 &     ''       &     ''       & 182\\
{\em Swift} & UVOT & $u$    & 00080730001 &     ''       &     ''       & 182\\
{\em Swift} & UVOT & $uvw1$ & 00080730001 &     ''       &     ''       & 364\\
{\em Swift} & UVOT & $uvm2$ & 00080730001 &     ''       &     ''       & 591\\
{\em Swift} & UVOT & $uvw2$ & 00080730001 &     ''       &     ''       & 731\\ \hline
\multicolumn{7}{c}{X-ray}\\ \hline
{\em Swift} & XRT & 0.5--10\,keV & 00080730001 & Apr 5, 0.4 h & Apr 5, 5.4 h & 2,372\\
{\em Suzaku} & XIS0/1/3 & 1.2--12\,keV & 409051010 & Apr 3, 17.65 h & Apr 5, 10.69 h & 59,711\\
{\em NuSTAR} & FPMA/B & 3--79\,keV & 80002021003 & Apr 4, 21.35 h & Apr 5, 12.69 h & 61,038\\
{\em Suzaku} & HXD/PIN & 13--65\,keV & 409051010 & Apr 3, 17.65 h & Apr 5, 10.69 h & 50,434\\
{\em Suzaku} & HXD/GSO & 50--240\,keV & 409051010 & Apr 3, 17.65 h & Apr 5, 10.69 h & 50,434\\ \hline
\end{tabular}
\end{center}
\end{minipage}
\end{table}

\begin{table}
\caption{Log of Kanata/HONIR Observations\label{tab:obs_kanata}}
\begin{minipage}{\linewidth}
\begin{center}
\footnotesize
\begin{tabular}{cccc} \hline \hline
Epoch  &  Filter       & MJD          & Exposure (s)\\ 
\hline\hline
1      &  $V$          & 56749.7221   & 136\\
1      &  $J$          & 56749.7232   & 120\\
1      &  $K_{\rm s}$   & 56749.7302   & 60\\
1      &  $B$          & 56749.7405   & 75\\\hline
3      &  $J$          & 56751.7170   & 120\\
3      &  $V$          & 56751.7173   & 136\\\hline
5      &  $J$          & 56754.8052   & 120\\
5      &  $V$          & 56754.8059   & 136\\\hline
\end{tabular}
\end{center}
\end{minipage}
\end{table}

\begin{table}
\caption{Log of SLT/U42 Observations\label{tab:obs_slt}}
\begin{minipage}{\linewidth}
\begin{center}
\scriptsize
\begin{tabular}{ccccc} \hline \hline
       &  Times of $g^{\prime}$ & Times of $r^{\prime}$ & Times of $i^{\prime}$ & Times of $z^{\prime}$\\
       &  Exposures           & Exposures           & Exposures           & Exposures\\
Epoch  &  (MJD-56750)         & (MJD-56750)         & (MJD-56750)         & (MJD-56750)\\
\hline\hline
2      & 0.3789 & 0.3806 & 0.3822 & 0.3839\\
2      & 0.3856 & 0.3872 & 0.3889 & 0.3905\\
2      & 0.3922 & 0.3938 & 0.3955 & 0.3971\\
2      & 0.3987 & 0.4004 & 0.4020 & 0.4037\\
2      & 0.4054 & 0.4070 & 0.4087 & 0.4103\\
2      & 0.4120 & 0.4136 & 0.4153 & 0.4169\\
2      & 0.4186 & 0.4202 & 0.4219 & 0.4235\\
2      & 0.4252 & 0.4268 & 0.4285 & 0.4301\\
2      & 0.4318 & 0.4334 & 0.4351 & 0.4367\\
2      & 0.4384 & 0.4400 & 0.4417 & 0.4433\\
2      & 0.4450 & 0.4466 & 0.4483 & 0.4499\\
2      & 0.4516 & 0.4532 & 0.4549 & 0.4566\\ \hline
3      & 1.7601 & 1.7578 & 1.7624 & 1.7647\\
3      & 1.7695 & 1.7672 & 1.7718 & 1.7741\\
3      & 1.7789 & 1.7766 & 1.7812 & 1.7835\\
3      & 1.7883 & 1.7860 & 1.7906 & 1.7929\\
3      & 1.7977 & 1.7954 & 1.8000 & 1.8023\\
3      & 1.8071 & 1.8048 & 1.8094 & 1.8117\\
3      & 1.8165 & 1.8142 & 1.8188 & 1.8211\\
3      & 1.8259 & 1.8236 & 1.8282 & 1.8305\\
3      & 1.8354 & 1.8330 & 1.8377 & 1.8400\\
3      & 1.8450 & 1.8428 & 1.8474 & 1.8498\\
3      & 1.8545 & 1.8523 & 1.8569 & 1.8592\\
3      & 1.8641 & 1.8617 & 1.8664 & ---\\ \hline
4      & 2.3799 & 2.3816 & 2.3833 & 2.3849\\
4      & 2.3866 & 2.3882 & 2.3899 & 2.3915\\
4      & 2.3932 & 2.3948 & 2.3965 & 2.3981\\
4      & 2.3998 & 2.4014 & 2.4031 & 2.4047\\
4      & 2.4064 & 2.4081 & 2.4097 & 2.4114\\
4      & 2.4130 & 2.4147 & 2.4163 & 2.4180\\
4      & 2.4196 & 2.4213 & 2.4229 & 2.4245\\
4      & 2.4262 & 2.4279 & 2.4295 & 2.4311\\
4      & 2.4328 & 2.4345 & 2.4361 & 2.4378\\
4      & 2.4394 & 2.4411 & 2.4427 & 2.4444\\
4      & 2.4460 & 2.4477 & 2.4493 & 2.4510\\
4      & 2.4526 & 2.4543 & 2.4560 & 2.4576\\ \hline
\end{tabular}
\end{center}
\end{minipage}
\end{table}

\begin{table}
\caption{Parameters for X-ray Spectral Fits\label{tab:parameters_x}}
\begin{minipage}{\linewidth}
\begin{center}
\footnotesize
\begin{tabular}{ccccccc} \hline \hline
Parameter &  Units & {\ttfamily pegpwrlw} & {\ttfamily highecut*} & {\ttfamily highecut*} & {\ttfamily highecut*} & {\ttfamily highecut*}\\
          &        &                      & {\ttfamily pegpwrlw}  & {\ttfamily pegpwrlw+} & {\ttfamily pegpwrlw+} & {\ttfamily pegpwrlw+}\\
          &        &                      &                       & {\ttfamily diskbb}    & {\ttfamily reflionx}  & {\ttfamily diskbb+}  \\
          &        &                      &                       &                       &                       & {\ttfamily reflionx}\\ 
\hline\hline
$N_{\rm H}$ 
& $10^{21}$\,cm$^{-2}$ 
                   & $2.01\pm 0.05$       & $2.01\pm 0.06$        & $2.35\pm 0.11$        & $2.60\pm 0.09$   & $2.51\pm 0.13$\\
$\Gamma$ 
& Photon index     & $1.722\pm 0.003$     & $1.721\pm 0.003$      & $1.699\pm 0.005$      & $1.774\pm 0.006$ & $1.738\pm 0.015$\\
Flux\footnote{unabsorbed 2--10\,keV, power-law only}
& $10^{-12}$\,erg\,cm$^{-2}$\,s$^{-1}$
                   & $253\pm 3$           & $253\pm 3$            & $253\pm 4$            & $254\pm 3$       & $253\pm 4$\\
$E_{\rm cut}$ & keV  & ---                  & $66^{+16}_{-9}$        & $60^{+12}_{-17}$       & $68^{+34}_{-15}$  & $67^{+20}_{-12}$\\
$E_{\rm fold}$ & keV & ---                  & $217^{+89}_{-72}$      & $209^{+568}_{-65}$     & $>$411           & $261^{+218}_{-95}$\\
$kT_{\rm in}$ & keV & ---                  & ---                   & $0.67^{+0.03}_{-0.05}$  & ---              & $0.67\pm 0.06$\\
$N_{\rm diskbb}$  & --- & ---               & ---                   & $3.8^{+1.5}_{-1.1}$     & ---              & $2.1^{+1.7}_{-1.1}$\\
$\xi$ & erg\,cm\,s$^{-1}$ & ---           & ---                   & ---                    & $<$5.3           & $<$11\\
Fe/solar & --- & ---                     & ---                   & ---                    & $0.28\pm 0.08$    & $0.47^{+0.21}_{-0.15}$\\
$N_{\rm refl}$ & $10^{-4}$ & ---     & ---                  & ---                    & $4.1^{+0.5}_{-3.3}$ & $2.2^{+1.0}_{-2.0}$\\
$\Omega/2\pi$ & --- & ---                & ---                   & ---                    & 0.20               & 0.12\\ \hline
$C_{\rm XRT}$ & ---  & 1.0                & 1.0                   & 1.0                    & 1.0                & 1.0\\
$C_{\rm XIS03}$ & --- & $1.06\pm 0.01$    & $1.06\pm 0.01$        & $1.04\pm 0.01$         & $1.04\pm 0.01$     & $1.04\pm 0.01$\\
$C_{\rm XIS1}$ & --- & $0.97\pm 0.01$     & $0.97\pm 0.01$        & $0.95\pm 0.01$         & $0.95\pm 0.01$     & $0.95\pm 0.01$\\
$C_{\rm FPMA}$ & --- & $1.08\pm 0.01$     & $1.08\pm 0.01$        & $1.07\pm 0.01$         & $1.06\pm 0.01$     & $1.06\pm 0.01$\\
$C_{\rm FPMB}$ & --- & $1.10\pm 0.01$     & $1.10\pm 0.01$        & $1.08\pm 0.01$         & $1.08\pm 0.01$     & $1.08\pm 0.01$\\
$C_{\rm PIN}$ & --- & $1.31\pm 0.02$      & $1.31\pm 0.02$        & $1.26\pm 0.02$         & $1.25\pm 0.01$     & $1.25\pm 0.02$\\
$C_{\rm GSO}$ & --- & $1.05\pm 0.06$      & $1.19\pm 0.08$        & $1.14\pm 0.09$         & $1.16\pm 0.07$     & $1.17\pm 0.08$\\ \hline
$\chi^{2}$/dof & --- & 2990/2143         & 2970/2141             & 2751/2139              & 2715/2138          & 2699/2136\\ \hline
\end{tabular}
\end{center}
\end{minipage}
\end{table}

\begin{table}
\caption{Parameters for Near-IR, Optical, UV, and X-ray Spectral Fits\label{tab:parameters2}}
\begin{minipage}{\linewidth}
\begin{center}
\scriptsize
\begin{tabular}{ccccc} \hline \hline
Parameter &  Units      & {\ttfamily diskir}    & {\ttfamily diskir+}   & {\ttfamily diskir+}\\
          &             &                       & {\ttfamily highecut*} & {\ttfamily reflionx}\\
          &             &                       & {\ttfamily pegpwrlw}  &                     \\
          &             &                       &                       &                     \\ \hline\hline
$E(B-V)$  &  ---        & 0.45\footnote{Fixed.} & 0.45$^{a}$            & 0.45$^{a}$\\
$N_{\rm H}$ & $10^{21}$\,cm$^{-2}$ 
                        & $2.08\pm 0.05$        & $2.83\pm 0.10$        & $2.60\pm 0.08$\\ \hline
\multicolumn{5}{c}{\ttfamily diskir}\\ \hline
$kT_{\rm in}$ & eV       & $12^{+8}_{-5}$         & $29^{+17}_{-7}$        & $29\pm 5$\\
$N_{\rm diskbb}$ & $10^{7}$    
                       & $87^{+607}_{-36}$       & $8.2^{+72}_{-6.1}$     & $5.5^{+18}_{-3.1}$\\
$\Gamma$ & Photon index  
                       & $1.734\pm 0.003$       & $1.90\pm 0.07$        & $1.777\pm 0.006$\\
$kT_{\rm e}$ & keV      & $>$60                  & $>$35                 & $>$429\\
$L_{c}/L_{d}$ & ---     & $4.2^{+2.4}_{-2.1}$     & $0.77\pm 0.17$        & $2.4\pm 0.6$\\
$f_{\rm in}$  & ---     & 0.1$^{a}$               & 0.1$^{a}$             & 0.1$^{a}$\\
$r_{\rm irr}$ & ---     & 1.1$^{a}$               & 1.1$^{a}$             & 1.1$^{a}$\\
$f_{\rm out}$ & ---     & 0.0$^{a}$               & 0.0$^{a}$             & 0.0$^{a}$\\
$\log{r_{\rm out}}$ & $\log{(R_{\rm out}/R_{\rm in})}$
                     & $1.83^{+0.06}_{-0.40}$    & $2.33^{+0.29}_{-0.11}$   & $2.59^{+0.34}_{-0.14}$\\ \hline
\multicolumn{5}{c}{\ttfamily highecut*pegpwrlw}\\ \hline
$\Gamma$ & 2nd Photon index     
                       & ---                   & $1.33^{+0.08}_{-0.25}$   & ---\\
Flux\footnote{unabsorbed 2--10\,keV, power-law only}
& $10^{-12}$\,erg\,cm$^{-2}$\,s$^{-1}$
                        & ---                   & $68^{+35}_{-42}$       & ---\\
$E_{\rm cut}$ & keV      & ---                   & $21^{+2}_{-3}$         & ---\\
$E_{\rm fold}$ & keV     & ---                   & $151^{+63}_{-26}$       & ---\\ \hline
\multicolumn{5}{c}{\ttfamily reflionx}\\ \hline
$\xi$ & erg\,cm\,s$^{-1}$ & ---                 & ---                    & $5.0^{+4.4}_{-2.2}$\\
Fe/solar & ---         & ---                   & ---                    &  $0.33\pm 0.09$\\
$E_{\rm fold}$ & keV     & ---                   & ---                    & $>$507\\
$N_{\rm refl}$ & $10^{-4}$ & ---           & ---                   & $0.78^{+0.22}_{-0.32}$\\
$\Omega/2\pi$ & ---     & ---                  & ---                    & 0.20\\ \hline
$C_{\rm XRT}$ & ---      & 1.0                  & 1.0                    & 1.0\\
$C_{\rm XIS03}$ & ---    & $1.06\pm 0.02$        & $1.04\pm 0.01$         & $1.04\pm 0.01$\\
$C_{\rm XIS1}$ & ---     & $0.97\pm 0.01$        & $0.94\pm 0.01$         & $0.95\pm 0.01$\\
$C_{\rm FPMA}$ & ---     & $1.08\pm 0.01$        & $1.06\pm 0.01$         & $1.07\pm 0.01$\\
$C_{\rm FPMB}$ & ---     & $1.09\pm 0.01$        & $1.07\pm 0.01$         & $1.08\pm 0.01$\\
$C_{\rm PIN}$ & ---      & $1.30\pm 0.02$        & $1.25\pm 0.02$         & $1.26\pm 0.02$\\
$C_{\rm GSO}$ & ---      & $1.17^{+0.06}_{-0.03}$   & $1.16\pm 0.07$         & $1.23\pm 0.07$\\ \hline
$\chi^{2}$/dof & ---    & 2963/2152            & 2765/2148              & 2769/2148\\ \hline
\end{tabular}
\end{center}
\end{minipage}
\end{table}

\begin{table}
\caption{Parameters for SED Fits\label{tab:parameters_sed}}
\begin{minipage}{\linewidth}
\begin{center}
\footnotesize
\begin{tabular}{ccc} \hline \hline
Parameter &  Units                             & Value\\ \hline\hline
$E(B-V)$  &  ---                               & 0.45\footnote{Fixed.}\\
$N_{\rm H}$ & $10^{21}$\,cm$^{-2}$          & $2.84\pm 0.11$\\ \hline
\multicolumn{3}{c}{\ttfamily diskir}\\ \hline
$kT_{\rm in}$ & eV                               & $28^{+21}_{-11}$\\
$N_{\rm diskbb}$ & $10^{7}$                 & $9.0^{+89}_{-1.2}$\\
$\Gamma$ & Photon index                        & $1.90\pm 0.02$\\
$kT_{\rm e}$ & keV                               & $>$33\\
$L_{c}/L_{d}$ & ---                              & $0.82^{+0.04}_{-0.19}$\\
$f_{\rm in}$  & ---                              & 0.1$^{a}$\\
$r_{\rm irr}$ & ---                              & 1.1$^{a}$\\
$f_{\rm out}$ & ---                              & 0.0$^{a}$\\
$\log{r_{\rm out}}$ & $\log{(R_{\rm out}/R_{\rm in})}$ & $2.31^{+0.06}_{-0.04}$\\ \hline
\multicolumn{3}{c}{\ttfamily highecut*bknpower}\\ \hline
$\Gamma_{1}$ & Below $E_{\rm break}$              & 0.7$^{a}$\\
$\Gamma_{2}$ & Above $E_{\rm break}$              & $1.33^{+0.10}_{-0.13}$\\
$E_{\rm break}$ & $10^{-6}$\,keV           & 0.1--15\\
$\nu_{\rm break}$ & Hz                           & $2.4\times 10^{10}$--$3.6\times 10^{12}$\\
Normalization & ph\,cm$^{-2}$\,s$^{-1}$\,keV$^{-1}$ at 1\,keV & $64\pm 2$\\
$E_{\rm cut}$ & keV                             & $20\pm 3$\\
$E_{\rm fold}$ & keV                            & $142^{+110}_{-25}$\\ \hline
$C_{\rm XRT}$ & ---                             & 1.0\\
$C_{\rm XIS03}$ & ---                           & $1.04\pm 0.01$\\
$C_{\rm XIS1}$ & ---                            & $0.94\pm 0.01$\\
$C_{\rm FPMA}$ & ---                            & $1.06\pm 0.01$\\
$C_{\rm FPMB}$ & ---                            & $1.07\pm 0.01$\\
$C_{\rm PIN}$ & ---                             & $1.25\pm 0.02$\\
$C_{\rm GSO}$ & ---                             & $1.16\pm 0.07$\\ \hline
$\chi^{2}$/dof & ---                          & 2768/2156\\ \hline
\end{tabular}
\end{center}
\end{minipage}
\end{table}

\begin{table}
\caption{Parameters for Power Spectrum Fits\label{tab:parameters_power}}
\begin{minipage}{\linewidth}
\begin{center}
\footnotesize
\begin{tabular}{ccc} \hline \hline
Parameter &  Units & Value\\ \hline \hline
\multicolumn{3}{c}{{\em Suzaku} (Lorentzian plus power-law)}\\ \hline
$\nu_{\rm max}$ & Hz & $0.110\pm 0.003$\\
$rms_{\rm Lor}$ & --- & 27.3\%$\pm 0.2$\%\\
Power-law index & --- & $2.12\pm 0.05$\\
$rms_{\rm pl}$ & 0.0001-1\,Hz & 4.4\%$\pm 1.4$\%\\
$\chi^{2}$/dof & --- & 284/191\\ \hline
\multicolumn{3}{c}{{\em Swift} (Lorentzian)}\\ \hline
$\nu_{\rm max}$ & Hz & $0.16\pm 0.04$\\
$rms_{\rm Lor}$ & --- & 22\%$\pm 2$\%\\
$\chi^{2}$/dof & --- & 51/42\\ \hline
\end{tabular}
\end{center}
\end{minipage}
\end{table}

\end{document}